\documentclass[aps,prb,twocolumn,showpacs,groupedaddress,floatfix]{revtex4}
\pdfoutput=1
\usepackage{graphicx}

\begin{document}

\title{Tuning the exchange interaction by electric field in laterally coupled quantum dots}
\author{Artur Kwa\'sniowski}
\author{Janusz Adamowski}\email[]{adamowski@ftj.agh.edu.pl}
\affiliation{Faculty of Physics and Applied Computer Science, AGH University of Science and
Technology, al. Mickiewicza 30, 30-059 Krak\'ow, Poland}

\date{\today}
\begin{abstract}
The effect of external electric field on the exchange interaction
has been studied by an exact diagonalization method for two
electrons in laterally coupled quantum dots (QD's). We have
performed a systematic study of several nanodevices that contain
two gate-defined QD's with different shapes and sizes located
between source and drain contacts.  The confinement potential is
modelled by two potential wells with a variable range and
softness. In all the considered nanodevices, the overall
dependence of exchange energy $J$ on electric field $F$ is
similar, i.e., for low fields $J$ increases with increasing $F$,
for intermediate fields $J$ reaches a maximum, and rapidly falls
down to zero if $F$ exceeds a certain critical value. However, the
$J(F)$ dependence shows characteristic properties that depend on
the nanodevice geometry. We have found that the low- and
intermediate-field behaviour can be accurately parametrized by
linear function $J(F) = \alpha F + \beta$, where $\alpha$ is
independent of the nanodevice geometry and softness of the
confinement potential. We have shown that the linear $J(F)$
dependence appears only if the tunnel coupling between the QD's is
weak, i.e., the interdot separation is sufficiently large. The
$J(F)$ dependence becomes non-linear for the strong interdot
tunnel coupling. If the QD located near the contact, to which the
higher voltage is applied, possesses the elliptic shape and is
larger than the other QD, the $J(F)$ dependence shows a plateau in
a broad electric-field regime.   The linearity and rapid jumps of
$J(F)$ as well as the existence of the plateau can be applied to
tune the exchange interaction by changing the external electric
field.
\end{abstract}


\maketitle

\section{Introduction}

Exchange interaction is one of the most characteristic quantum
effects in many-electron systems.  In natural atoms and molecules,
it leads to a singlet-triplet splitting of energy levels and
is responsible for a binding of atoms into a molecule.
In solids, it gives rise to a covalent bonding of elemental semiconductors
and ferromagnetic properties of metals.
In the absence of external fields, the exchange
interaction in natural atomic systems is determined by the charges
of nuclei and the number of electrons and is essentially fixed.
Man-made solid-state analogs of atomic systems, namely, quantum dots (QD's),
called also artificial atoms, and coupled QD's (artificial molecules)
can be fabricated in various designed shapes and sizes.  One can also change the depth and range of the
potential confining the electrons.  This gives us a unique
opportunity of engineering the quantum states of electrons
confined in the QD's and tuning the exchange interaction.
The exchange interaction between electrons in QD's
has been proposed as an effective mechanism for
changing the electron spin, i.e., for performing the quantum logic operations
on spin qubits.\cite{loss98,burk99}  This mechanism seems to be very promising
in quantum information processing with the solid-state nanodevices.

The investigation of the exchange interaction in the QD-based nanodevices
allows us to elaborate the methods of controlling and tuning this interaction,
which in turn leads to the controlled manipulation of the electron
spin qubits.\cite{loss98,burk99,burk00,petta}
Recently, the exchange-interaction induced spin swap operations in coupled QD's have been simulated
by a direct solution of a time-dependent Schr\"{o}dinger equation.\cite{mosk07}
The coherent manipulation of spin qubits in lateral QD's has been
studied experimentally by Petta et al.,\cite{petta} Elzerman et al.,\cite{elz03,elz04} and Hayashi et al.\cite{hay}
Hatano et al.\cite{hat05} determined the tunnel and exchange couplings
in laterally coupled vertical QD's. The quantum logic operations can also be performed in the
nanowire double QD's.\cite{fuhrer}

The exchange energy is defined as
\begin{equation}
J=E_T-E_S \;,
\label{exch}
\end{equation}
where $E_T$ and $E_S$ are the lowest triplet and singlet energy levels, respectively.
In physics of solid-state nanodevices, exchange energy (\ref{exch})  plays a two-fold role:
(i) it determines the strength of the exchange interaction (Heisenberg interaction),
which can swap the spin qubits; \cite{loss98,burk99,burk00,mosk07}
(ii) $J$ can be treated as the exchange splitting (singlet-triplet splitting) that allows us to distinguish the different
spin states of the electron system.\cite{petta}  In both the cases,
we should know how to tune $J$ by applying the external fields.

In the coupled QD system, the exchange energy was calculated by several
groups.\cite{burk99,burk00,naga99,wens,yan,jeff,harju,marlo,sz04,bell,dyb,kim,zh06,zh07,zh08,zh08a,ped,nowak,ka08}
The effect of an external magnetic field was investigated in papers.\cite{burk99,burk00,bell,sz04,zh06,zh07}
The magnetic field, applied perpendicular to the plane of the electron movement,
lowers the energy of the triplet state, which decreases the exchange splitting.
The asymmetry of the QD's gives rise to an enhancement of the exchange interaction.\cite{sz04,zh07}
The size effects in the exchange coupling were studied in papers.\cite{ped,zh08a,ka08}
The increasing size of the coupled-QD system leads to the decrease of the exchange energy.\cite{ped,zh08a,ka08}
For the two identical elliptic QD's Zhang et al.\cite{zh07,zh08} calculated the exchange energy
as a function of aspect ratio $r=R_y/R_x$, where $R_x (R_y)$ is the extension of the QD in the $x (y)$ direction.
They obtained the increase of $J$ with increasing $r$ for $r \leq 1.5$\cite{zh07}
and the sharp variation of $J$ as a function of interdot detuning for $r \geq 3$.\cite{zh08}
The influence of an external electric field on the exchange energy was studied
in papers\cite{burk00,nowak} for vertically coupled self-assembled QD's.
Burkard et al.\cite{burk00} assumed the harmonic confinement potential
and calculated the exchange energy for the vertically coupled QD's using the Heitler-London and Hund-Mulliken techniques.
These results\cite{burk00}  show the monotonic decrease of the exchange energy with the increasing in-plane electric
field.  Pedersen et al.\cite{ped} calculated the exchange energy
for the harmonic double-dot confinement potential using the numerically exact approach and
showed the failure of standard approximations (i.e., Heitler-London, Hund-Mulliken, Hubbard)
even for simple model systems.
In the present paper, we have elaborated a numerical procedure that provides accurate results
for the lowest-energy states of the two electrons in laterally coupled QD's.
Using this method, we have performed a systematic study of QD nanodevices with different geometry
and confinement potential profile.  In this study,
we have applied the confinement potentials with a different softness,
i.e., we have taken into account a variable smoothness of the QD interface.\cite{mlin}
We have investigated a large class of realistic potentials with the finite depth:
from the soft Gaussian potential to the hard rectangular-like potential.\cite{ka08,ciu}

A high fidelity of quantum logic operations on spin qubits can be achieved if the exchange interaction is
possibly strong.  This leads to the problem of designing such a nanodevice, in which the exchange interaction is maximal.
A possibility of tuning the exchange interaction with the help of external fields is another important issue
of quantum computing in solid-state nanodevices. In the present paper, we focus on
the electric-field induced tuning of the exchange interaction in laterally coupled QD's.
We note that the electronic properties of the electrostatically gated QD's \cite{lis03,hand}
can be tuned by changing the voltages applied to the gates, which changes the potential confining
the electrons in the nanodevice.  This leads to another method of tuning the exchange interaction
by changing the gate voltages.
In the present paper, we have also investigated this method by studying the effect
of the variable range and softness of the confinement potential on the exchange energy.
In the gate-defined QD's,\cite{elz03,elz04} the range and softness of the confinement potential are determined
by the voltages applied to the gates.\cite{lis03,hand,lis08}

The present paper is organized as follows: in Section 2, we briefly describe the
theoretical model and the computational method used in the calculations. Section 3 contains
the numerical results, Section 4 -- discussion, and Section 5 --
conclusions and summary.  The details of the computational approach are presented
in Appendix.

\section{Theory}

We study the system of two electrons confined in two laterally
coupled QD's and subjected to a static homogeneous electric field.
The lateral QD's are usually created in a quasi-two-dimensional
electron gas by applying the suitably chosen voltages to the
gates, which are placed on the surface of the nanodevice above the plane, in which
the electrons move.\cite{elz03,elz04,petta}  Therefore, we have
assumed the two-dimensional (2D) motion of the electrons. Figure 1
displays (a) the geometry of the nanostructure and (b) the
confinement potential profile in the $x$ direction. We assume that
the potential energy of the electron in the single QD is described
by the power-exponential function \cite{ciu}
\begin{equation}
U_{\mu}(\mathbf{r})=-U_{0\mu}e^{-[(x-x_{0\mu})^2/R_{\mu x}^2
+(y-y_{0\mu})^2/R_{\mu y}^2]^{p/2}} \;,
\label{conf1}
\end{equation}
where index $\mu$ labels the QD's ($\mu=l$ and $r$ for the left and
right QD, respectively),  $U_{0\mu}$ is the potential well depth ($U_{0\mu} >0$),
$\mathbf{r}=(x,y)$, $\mathbf{r}_{0\mu}=(x_{0\mu},y_{0\mu})$ is the
position of the QD center, and $R_{\mu x}$ ($R_{\mu y}$) is the
range of the confinement potential in the $x$ ($y$) direction,
i.e., it determines the extension of the QD in the corresponding direction.
Parameter $p$ ($p \geq 2$) describes the softness of the confinement potential at the QD boundaries, i.e.,
the smoothness of the QD interfaces.\cite{mlin}
For $p=2$ we deal with the soft Gaussian potential, while for $p
\geq 4$ the potential can be treated as ''hard'', in particular,
for $p \longrightarrow \infty$ potential energy (\ref{conf1}) takes on the
rectangular shape. Form (\ref{conf1}) of the confinement
potential energy allows us to model a large variety of QD's with a
different shape, size, and interface smoothness.

\begin{figure}
\includegraphics[width=0.6\linewidth,height=\linewidth,angle=270]{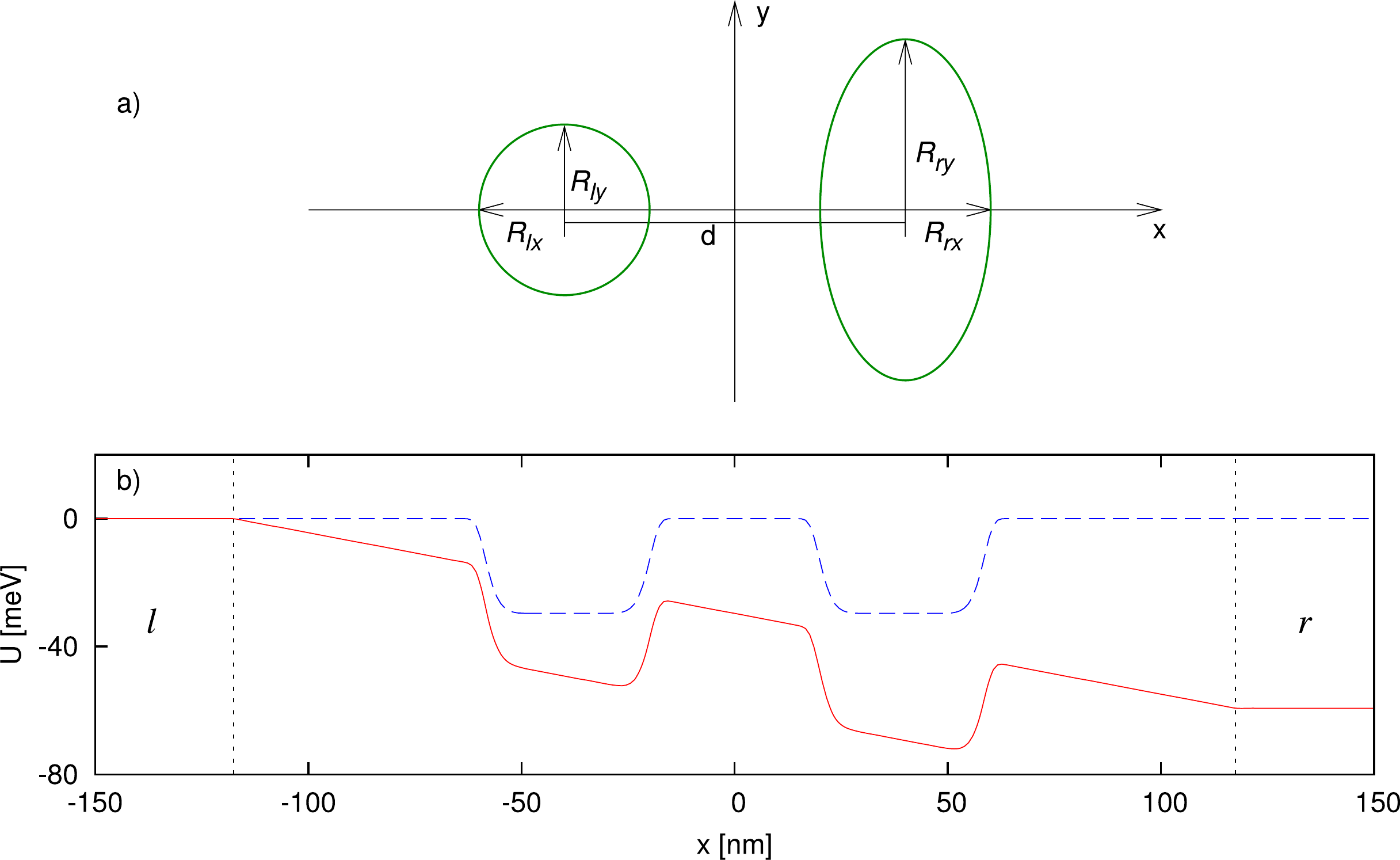}
\caption{\label{fig1} (Color online) (a) Schematic of the coupled QD's.
Solid (green) curves correspond to the extensions of the QD's.
(b) Profile of confinement potential energy $U$ plotted as a function of $x$ for $y=0$
with [solid (red) curve] and without [dashed (blue) curve] the external electric field.
The vertical lines show the boundaries of the left ($l$) and right ($r$) electrode.
The parameters of the nanodevice in panel (b):
$R_{lx} = R_{ly} =20$ nm, $R_{rx} = 20$ nm, $R_{ry} = 40$ nm, $d = 80$ nm, $p=10$, and $V=70$ mV.}
\end{figure}

For the coupled QD's the confinement potential is the sum of single QD confinement potentials (\ref{conf1})
\begin{equation}
U_{con\!f}(\mathbf{r}) = U_l(\mathbf{r})+U_r(\mathbf{r}) \;.
\label{conf2}
\end{equation}
The $x$ axis is directed along the straight line connecting the centers of both the QD's [cf. Fig. 1(a)].
Here, we take on $x_{0l}=-x_{0r}$ and $y_{0l}=y_{0r}=0$.
The QD's are separated by the potential barrier, i.e., distance $d$
between the QD centers is larger than $R_{lx}+R_{rx}$.

We investigate the nanodevice, which consists of the coupled QD's placed between
the left and right metal contacts [cf. Fig. 1(b)].
The electrodes are separated by a finite distance $L$ and a static external voltage $V$ is applied between them.
In the semiconductor region, the electrodes generate the homogeneous electric field $\mathbf{F}=(-F,0,0)$,
where $F= V/L$.
In the present calculations, distance $L$ between the boundaries of electrodes
is related to other geometric parameters as follows: $L \geq d+ 2(R_{lx}+R_{rx})$.
In electric field $\mathbf{F}$, each electron possesses additional potential energy $\Delta U(\mathbf{r})$
given by
\begin{eqnarray}
\Delta U(\mathbf{r}) = \left\{
\begin{array}{ll}
0 & \textrm{for $x<-L/2$} \;,\\
-eFx - eV/2 & \textrm{for $|x|\leq L/2$} \;,\\
-eV & \textrm{for $x>L/2$} \;.
\end{array} \right.
\label{DU}
\end{eqnarray}
In the present calculations, we measure the energy with respect to the electrochemical potential
$\mu_l$ of the left contact, i.e., we set $\mu_l=0$.
On the contrary to papers,\cite{zh06,nowak} in which the infinite range of electric field $\mathbf{F}$ is assumed,
we assume a more realistic space distribution of the electric field with the finite range.
Formula (\ref{DU}) gives the profile of the electron potential energy in the electric field created
by external voltage $V$, which -- in the real nanodevices --
is applied between the source and drain contacts separated by the finite distance.
Formulas (\ref{conf1}), (\ref{conf2}), and (\ref{DU}) set up a model of the nanodevice (Fig. 1),
which consists of the left ($l$) and right ($r$) metal electrodes, and
the semiconductor material, in which both the QD's are embedded.
The QD's are separated by the barrier potential region.
The total potential energy of the single electron is given by
\begin{equation}
U(\mathbf{r})=U_{con\!f}(\mathbf{r})+\Delta U(\mathbf{r}) \;.
\label{pot}
\end{equation}

In the effective mass approximation, the Hamiltonian of the two-electron system in the coupled QD's
reads
\begin{equation}
H = h_1 + h_2 + \frac{e^2}{4\pi\varepsilon_0\varepsilon_s r_{12}} \;,
\label{H_tot}
\end{equation}
where $h_j$ ($j=1,2$) is the one-electron Hamiltonian,
$\varepsilon_0$ is the electric permittivity of the vacuum,
$\varepsilon_s$ is the static relative electric permittivity of the semiconductor,
$r_{12}= |\mathbf{r}_1-\mathbf{r}_2|$ is the electron-electron distance,
and $\mathbf{r}_j$ is the position vector of the $j$th electron.
The one-electron Hamiltonian has the form
\begin{equation}
h_j = -\frac{\hbar^2}{2m_e}\nabla_j^2 + U(\mathbf{r}_j) \;,
\label{h1}
\end{equation}
where $m_e$ is the electron effective band mass.
We assume that the electron effective mass and the static electric
permittivity do not change across the QD boundaries.
This assumption is well satisfied for the GaAs-based electrostatic QD's.\cite{lis03,hand}

We solve the two-electron eigenvalue problem by a configuration interaction (CI) method,
which is performed in few steps.
First, we find one-electron orbital wave functions $\phi_{\nu}(x,y)$
using the expansion in a multicenter Gaussian basis (see Appendix).
In the second step, we transform the one-electron orbitals $\phi_{\nu}(x,y)$ into
the discrete representation $\phi_{\nu}^{mn}=\phi_{\nu}(x_m,y_n)$ on the two-dimensional grid $(x_m,y_n)$.
More details of this method are given in Appendix.
Augmenting the one-electron orbitals  by the eigenfunctions $\chi_{\sigma}$ of the $z$ component
of the electron spin we obtain one-electron spin-orbitals $\psi_{\nu\sigma}^{mn}$,
where $\nu$  is the set of orbital quantum numbers and $\sigma$ is the spin quantum number.
Spin-orbitals $\psi_{\nu\sigma}^{mn}$ are used to a construction of Slater determinants.
In the final step, we construct the two-electron wave function
as a linear combination of $N_S$ Slater determinants and solve the two-electron eigenvalue equation by the
exact diagonalisation.
All the potential energy matrix elements (including the electron-electron interaction energy)
have been calculated with a high precision by the numerical quadrature subroutines.
We have performed test calculations for $N_S$ = 64, 81, 100, 144, and  169
and obtained a good convergence for the lowest-energy levels.
A good compromise between the numerical accuracy and computer time has been found for $N_S = 81$;
therefore, the majority of calculations has been performed with  81 Slater determinants.
Finally, we calculate the lowest singlet ($E_S$) and triplet ($E_T$) energy levels,
and exchange interaction energy $J$ [Eq.~(\ref{exch})].
In the calculations, we have used the material parameters of GaAs, i.e.,
$\varepsilon_s = 12.4$ and $m_e = 0.067 m_{e0}$, where $m_{e0}$ is the free electron rest mass,
and fix the depth ($U_{0l}=U_{0r}= 30$ meV) of the confinement potential.
The present calculations have been performed for circular and elliptic QD's with aspect ratios\cite{zh07}
$r= 0.5$, 1, and 2.
The exchange energy has been calculated as a function of external electric field $F$ for different shapes,
sizes, and geometric configurations of the coupled QD's.

\section{Results}

\begin{figure}
\includegraphics[width=0.7\linewidth,height=\linewidth,angle=270]{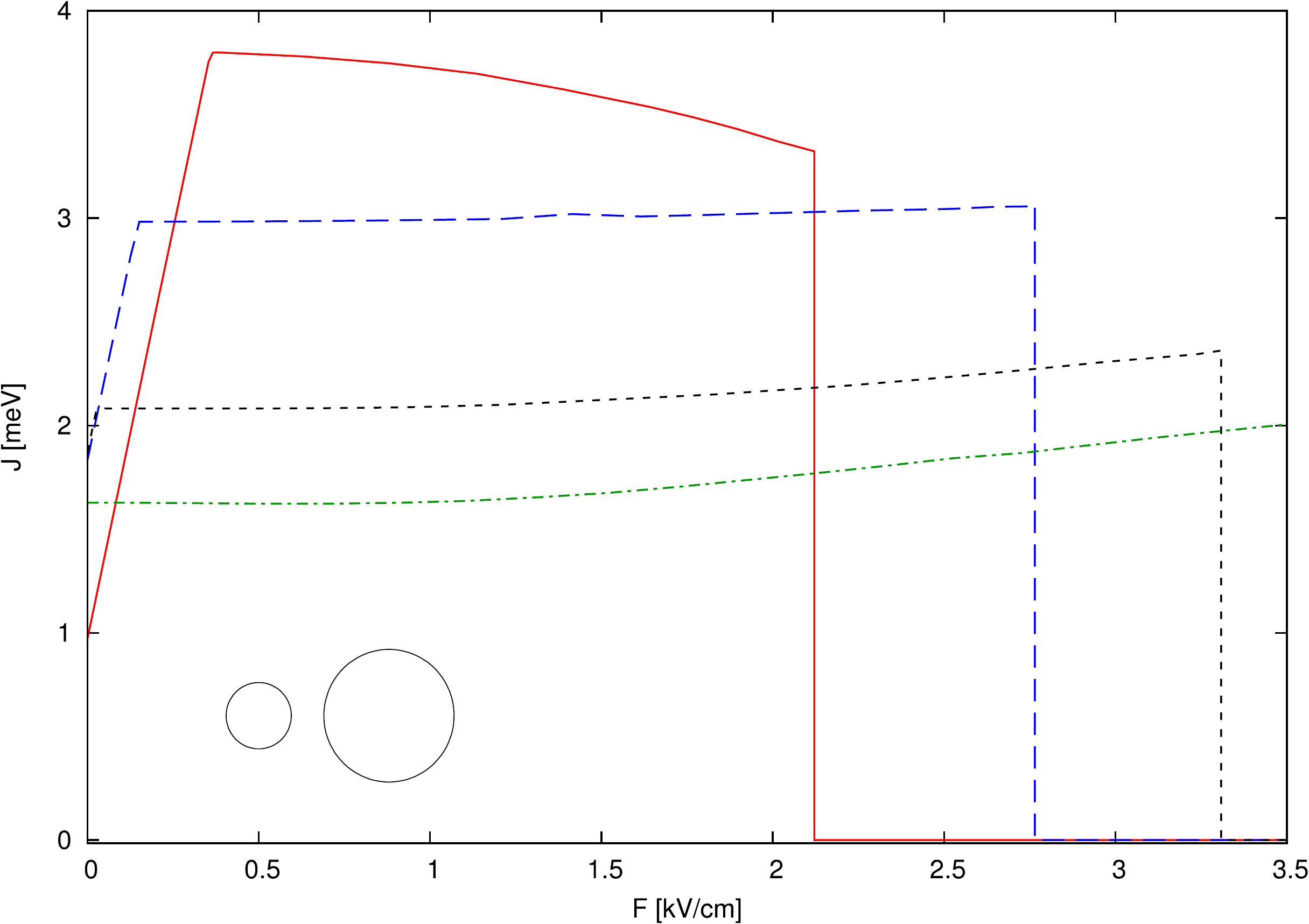}
\caption{\label{fig2} (Color online) Exchange energy $J$ as a function of electric field $F$
and softness parameter $p$ for two coupled circular QD's with $R_{lx}=R_{ly} = 20$ nm, $R_{rx}=R_{ry} = 40$ nm,
and $d = 80$ nm. Inset: schematic of the nanodevice.}
\end{figure}

Two lateral QD's form 16 geometric configurations that differ from
each other by their relative size (large/small QD), shape
(circular/elliptic QD), position with respect to the electrodes
(left/right QD), and orientation with respect to the electric
field (ellipse axis parallel/perpendicular to $\mathbf{F}$). In
this paper, we present the results for the four most
characteristic configurations (cf. insets of Figs. 2-5).  The
preliminary results for the two circular QD's with the same radius
have been presented in paper.\cite{pss}  Figures 2-4 show the
results for the nanodevices, in which the right QD is larger than
the left one. We remind that the left (right) QD is located near
the electrode with the higher (lower) potential energy of the
electron (cf. Fig. 1). These results (Figs. 2-4) have been
obtained for the left circular QD with fixed size, i.e.,
$R_{lx}=R_{ly} = 20$ nm, and for the different shapes and sizes of
the right QD: circular (Fig. 2), $y$-elongated elliptic (Fig. 3),
and $x$-elongated elliptic (Fig. 4). We have found that -- in
these nanodevices (cf. the insets of Figs. 2-4) -- the general
electric-field dependence of the exchange energy is similar. In
the low-field regime, the exchange energy takes on either small (Fig. 2)
or zero (Figs. 3 and 4) values, at higher fields, increases with
the electric field, and -- in the intermediate-field regime --
exhibits a cusp followed by a broad plateau region (Figs. 2 and
3), in which $J(F)$ takes on maximal values. At the sufficiently high
electric field, the $J(F)$ curve possesses the second cusp and
rapidly falls down to zero.  Figures 2-4 also show another general
property of the exchange energy: the maximal values $J_{max}$,
reached in the plateau region, increase with with decreasing $p$,
i.e., with the increasing softness of the confinement potential.
However, the detailed $J(F)$ dependence is different for each of
the nanodevice geometry considered.

\begin{figure}
\includegraphics[width=0.7\linewidth,height=\linewidth,angle=270]{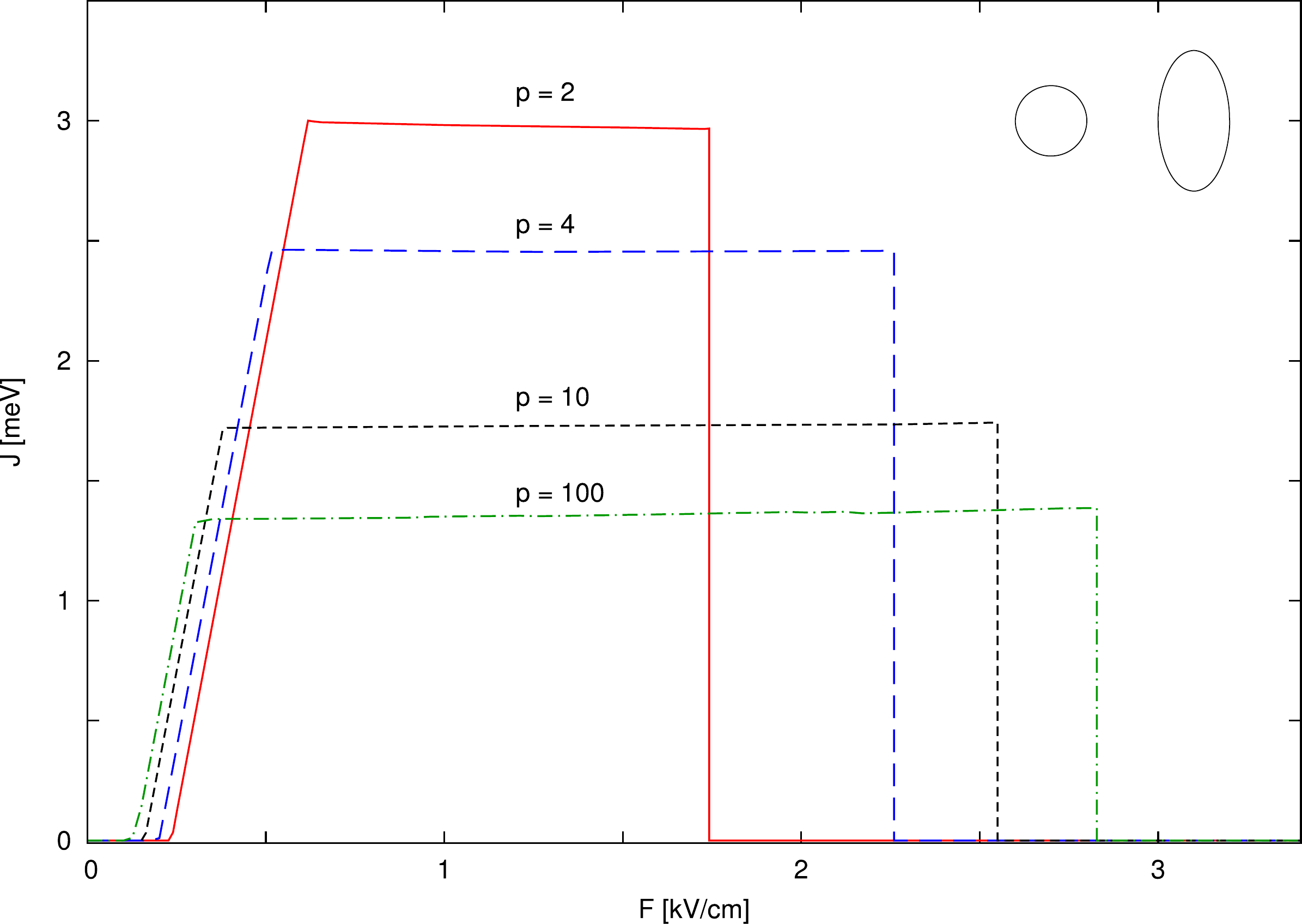}
\caption{\label{fig3} (Color online) Exchange energy $J$ as a function of electric field $F$
and softness parameter $p$ for coupled circular and elliptic QD's with $R_{lx}=R_{ly} = 20$ nm,
$R_{rx}=20$ nm, $R_{ry} = 40$ nm, and $d = 80$ nm.  Inset: schematic of the nanodevice.}
\end{figure}

\begin{figure}
\includegraphics[width=0.7\linewidth,height=\linewidth,angle=270]{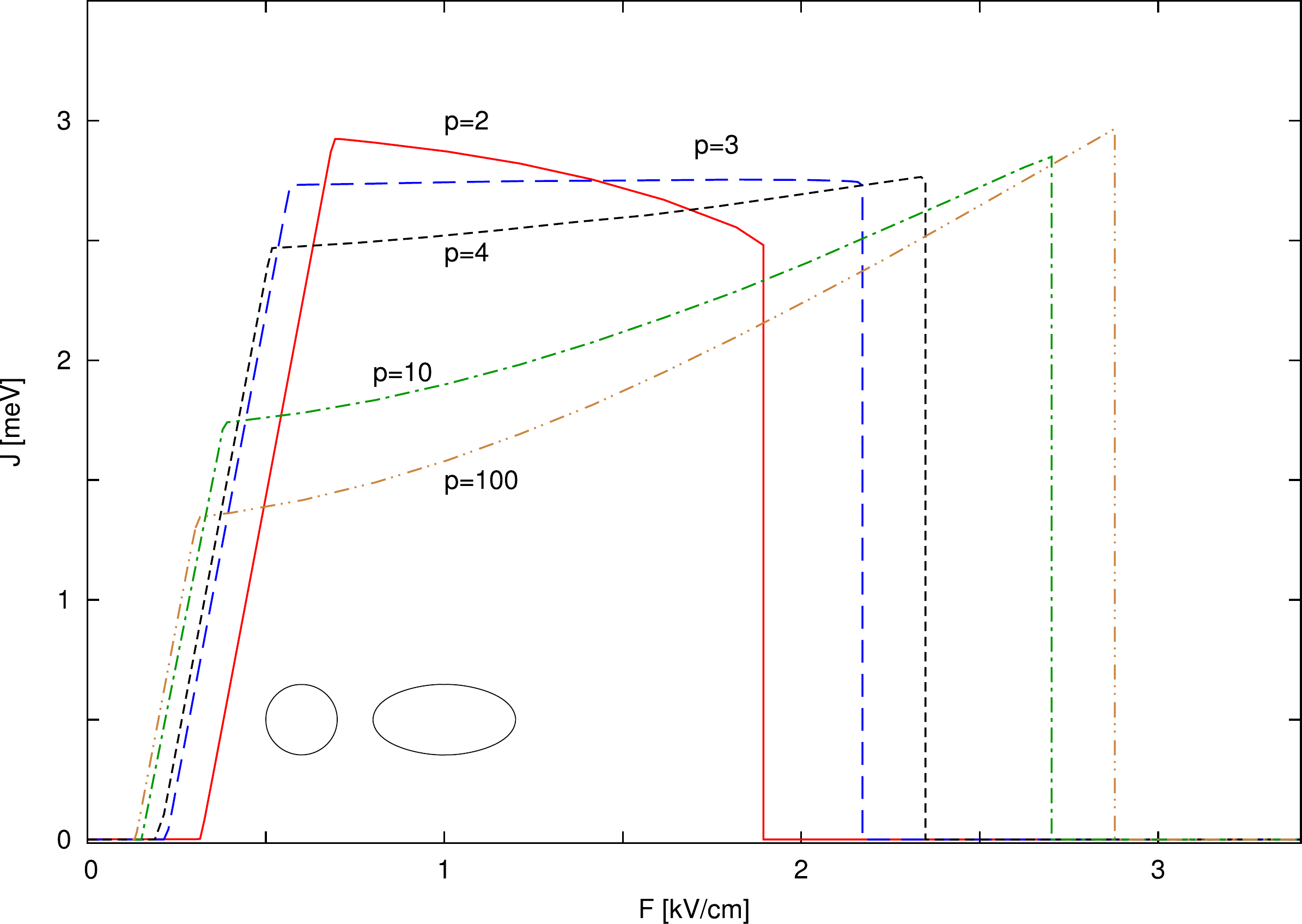}
\caption{\label{fig4} (Color online) Exchange energy $J$ as a function of electric field $F$
and softness parameter $p$ for coupled circular and elliptic QD's with $R_{lx}=R_{ly} = 20$ nm,
$R_{rx}=40$ nm, $R_{ry} = 20$ nm, and $d = 80$ nm.  Inset: schematic of the nanodevice.}
\end{figure}

The details of the low-field exchange energy behaviour are
different in the nanodevices with circular (Fig. 2) and elliptic
(Figs. 3-5) QD's. For the circular QD's (Fig. 2) the exchange
energy is non-zero in the absence of electric field. In the
nanodevice shown in the inset of Fig. 2, the electrons in the singlet state occupy
the right QD with a quite large probability already for $F=0$.
Even the weak electric field causes that both the electrons become
entirely localized in the right QD, i.e., the double QD system
starts to act as the single QD. If one of the QD's is elliptic
(Figs. 3-5), the exchange interaction vanishes in the low-field
regime, i.e., for $0 \leq F \leq F_{c0}$, becomes non-zero at $F =
F_{c0}$, and increases linearly with $F$ for $F_{c0} \leq F \leq
F_{c1}$. At $F = F_{c1}$ the $J(F)$ dependence exhibits the first
cusp, above which $J(F)$ is nearly constant (cf. the plateau
regions in Figs. 2 and 3) or changes slowly with the electric
field (cf. Fig. 4). The exchange energy reaches
maximal values $J_{max}$ for $F_{c1} \leq F \leq F_{c2}$, exhibits
the second cusp at $F=F_{c2}$, and rapidly vanishes for $F > F_{c2}$.
For the nanodevice with the larger $x$-elongated QD (Fig.
4) the exchange energy shows the variable behaviour in the
interval $F_{c1} \leq F \leq F_{c2}$ depending on the confinement
potential softness. According to Fig. 4, $J$ decreases with
increasing $F$ for $p=2$, is nearly constant for $p=3$, and
increases with increasing $F$ for $p \geq 4$. In the nanodevices
shown in the insets of Figs. 3 and 4, in the low-field regime, the
electrons are localized in different QD's, i.e., the overlap of
the corresponding one-electron wave functions vanishes, which
leads to the vanishing exchange interaction.
If the electric field exceeds the critical value $F_{c0}$,
the electrons in the singlet state are entirely localized in the right QD,
while the electrons in the triplet state become
more and more localized in the right QD.  This leads to
the increase of the exchange energy in the interval
$F_{c0} \leq F \leq F_{c1}$. The plateaus on the $J(F)$ dependence
(Figs. 2-4) result from the fact that both the electrons are
localized in the right QD and this localization is almost
unchanged in the interval $F_{c1} \leq F \leq F_{c2}$.  For the
sufficiently strong electric field the exchange interaction is equal to
zero since one of the electrons tunnels out of the QD system and
is absorbed in the right electron reservoir.

\begin{figure}
\includegraphics[width=0.7\linewidth,height=\linewidth,angle=270]{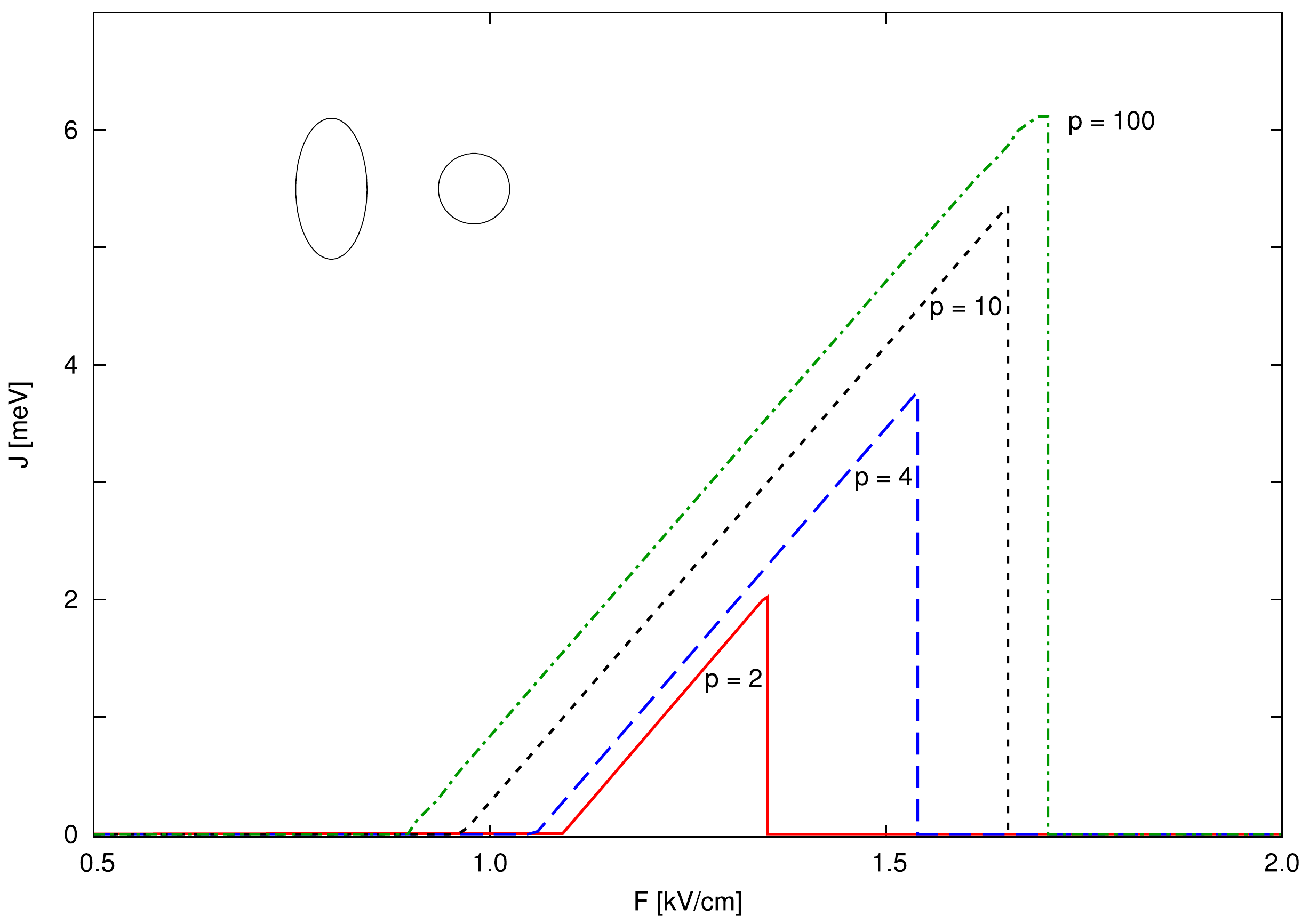}
\caption{\label{fig5} (Color online) Exchange energy $J$ as a function of electric field $F$
and softness parameter $p$ for coupled circular and elliptic QD's with $R_{lx}=20$ nm, $R_{ly} = 40$ nm,
$R_{rx}=R_{ry} = 20$ nm,
and $d = 80$ nm.  Inset: schematic of the nanodevice.}
\end{figure}

Figure 5 displays the results for the nanodevice with the left QD larger than the right one.
For $F \leq F_{c0}$ there is no exchange interaction.
If the electric field $F$ exceeds $F_{c0}$, the exchange energy becomes non-zero
and increases as a linear function of the electric field.
After reaching the maximum at $F=F_{c1}$, the exchange energy exhibits the sharp cusp and falls down to zero.
In the nanodevice shown in the inset of Fig. 5, the plateau region does not exist, which means that
$F_{c1} = F_{c2}$.

The results of Figs. 2-5 show that -- in the low- and intermediate-field regime --
the exchange energy is a linear function of the electric field and can be
parametrized as follows:
\begin{equation}
J(F)= \alpha F +\beta \;.
\label{linear}
\end{equation}
In the nanodevices depicted in the insets of Figs. 3-5,
the linear parametrization (\ref{linear}) is valid in the electric-field
interval $\Delta F_{linear} = F_{c1}-F_{c0}$.
Parameter $\beta$ depends on the softness of the confinement potential and the geometry of the nanodevice.
In general, $\beta$ increases with increasing $p$ and -- for the nanodevice shown in the inset of Fig. 5 --
takes on the values from $-8.56$ meV for $p=2$ to $-6.86$ meV for $p=100$.
The results of Figs. 3-5 show that parameter $\alpha$ is independent of the confinement potential softness
and the geometry of the nanodevice.
It takes on nearly constant value $\alpha \simeq 7.73$ [meV/(kV/cm)]
for all the nanodevices studied in the present work.
The physical interpretation and possible applications of the linear dependence
[Eq.~(\ref{linear})] will be discussed in Section 4.

The results presented in Figs. 2-5 can be explained if we consider the spatial localization of electrons.
It is convenient to illustrate the distribution of the electrons in the coupled QD's with the help of
the electron density $\varrho(\mathbf{r})$ defined as
\begin{equation}
\varrho(\mathbf{r})=\sum\limits_{j=1}^2 \langle \Psi | \delta(\mathbf{r}-\mathbf{r}_j) | \Psi \rangle \; ,
\label{rho}
\end{equation}
where $\Psi=\Psi(\mathbf{r}_1,\mathbf{r}_2)$ is the two-electron wave function.

\begin{figure}
\includegraphics[width=\linewidth,height=\linewidth,angle=270]{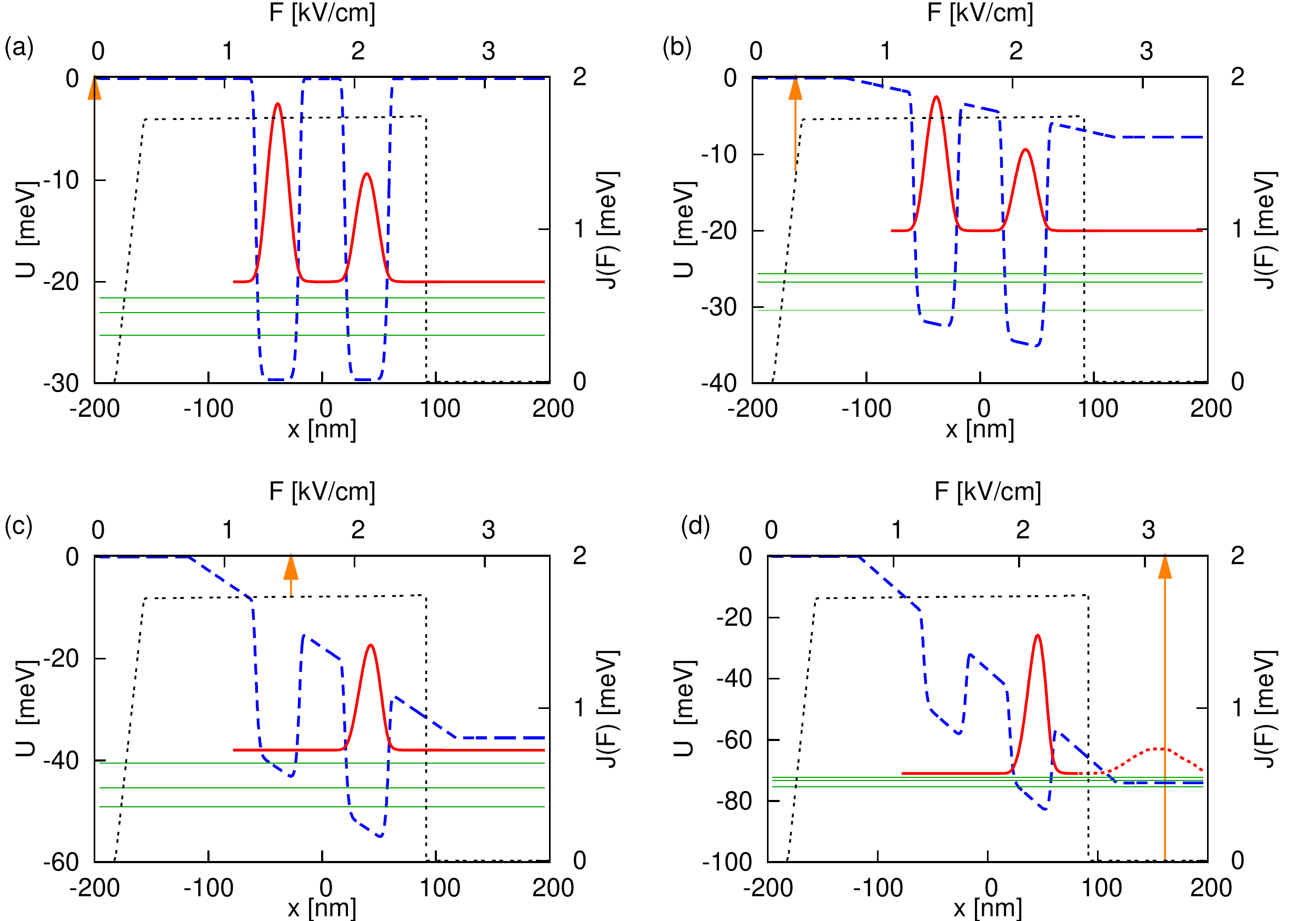}
\caption{\label{fig6} (Color online) Electron density for the triplet state
[solid (red) curve]
and confinement-potential profile [dashed (blue) curve] as functions of $x$ coordinate for $y=0$.
The horizontal (green) lines correspond to the three lowest one-electron energy levels.
Dotted curves show the exchange energy $J$ as a function of electric field $F$ (cf. Fig. 3).
The vertical (orange) arrows show the electric field $F$ for which the plots of electron density
and confinement potential are drawn.  In panel (d), the dotted (red) curve depicts
the electron density
of the electron just after the tunneling through the triangular barrier.
The left QD has the circular shape with $R_{lx}=R_{ly} = 20$ nm,
and the right QD has the elliptic shape with $R_{rx}=20$ nm and $R_{ry} = 40$ nm
(cf. inset of Fig. 3).  All the plots are drawn for $p=10$ and $d = 80$ nm.}
\end{figure}

\begin{figure}
\includegraphics[width=\linewidth,height=\linewidth,angle=270]{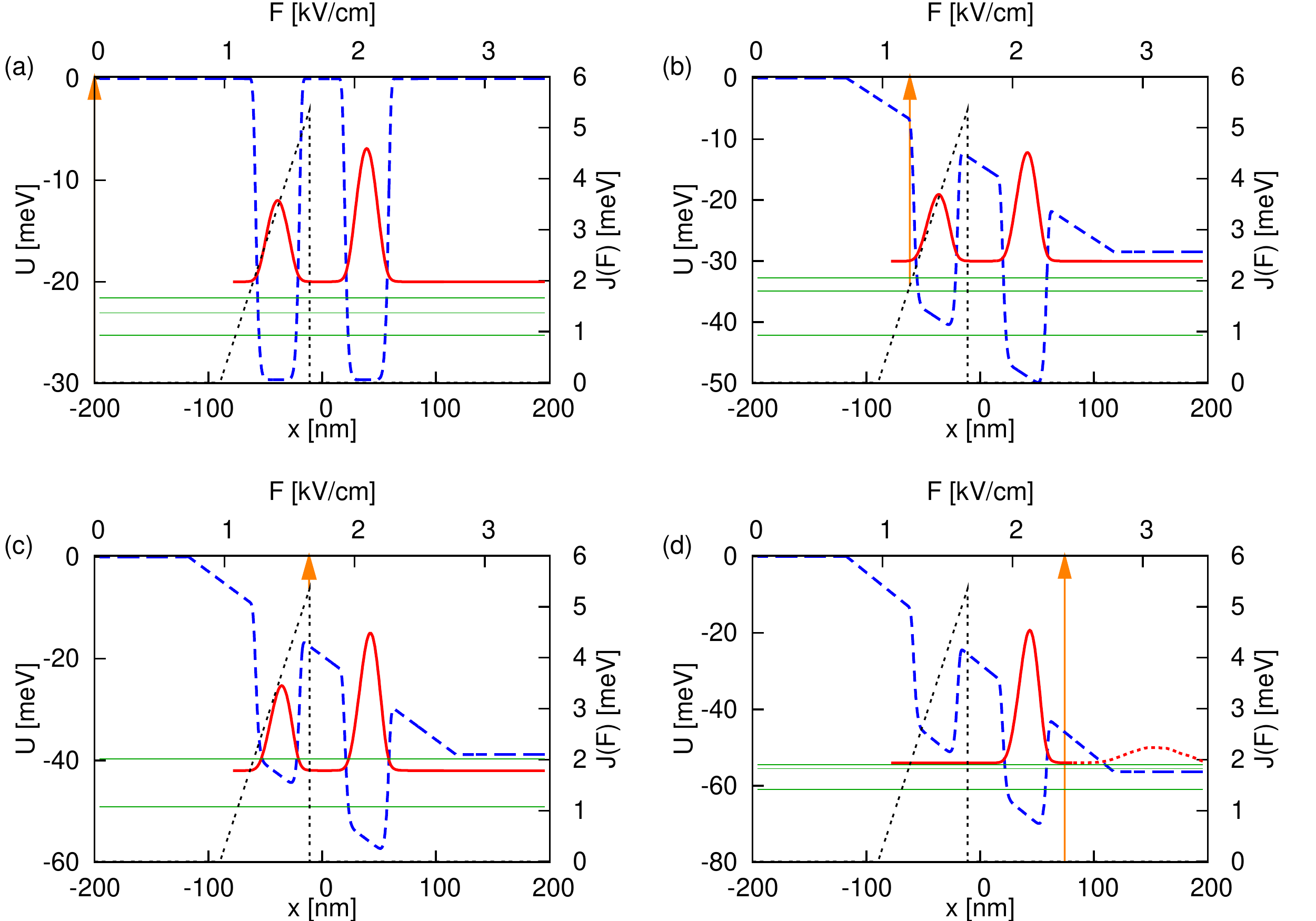}
\caption{\label{fig7} (Color online)
One-electron density for the triplet state [solid (red) curve]
and confinement-potential profile [dashed (blue) curve] as functions of $x$ coordinate for $y=0$.
The horizontal (green) lines correspond to the three lowest one-electron energy levels.
Dotted curves display the exchange energy $J$ as a function of electric field $F$ (cf. Fig. 5).
The vertical (orange) arrows show the electric field for which the plots of probability density
and confinement potential are drawn.
In panel (d), the dotted (red) curve depicts the probability density
of the electron just after the tunneling through the triangular barrier.
The left QD has the elliptic shape with $R_{lx}=20$ nm and $R_{ly} = 40$ nm,
the right QD has the circular shape with $R_{rx}=R_{ry} = 20$ nm
(cf. inset of Fig. 5).  All the plots are drawn for $p=10$ and $d = 80$ nm.}
\end{figure}

Solid (red) curves in Figs. 6 and 7 display the cross sections
of the electron density $\varrho(x,y)$
by the plane $y=0$ for the triplet state.
The corresponding profiles of the one-electron potential energy are plotted by dashed (blue) curves
in Figs. 6 and 7.
The horizontal (green) lines show the three lowest-energy levels $E_{\nu}$ of one-electron states, which
mostly contribute to the two-electron triplet wave function.
In Figs. 6 and 7, we have also presented the $J(F)$ dependencies
obtained for these nanodevices (cf. Figs. 3 and 5 for $p=10$).
The vertical (orange) arrows show the values of the electric field,
for which the plots of $\varrho$, $U$, and $E_{\nu}$ are drawn.
Figures 8 and 9 display the contours of the electron density on the $x-y$ plane
for the singlet and triplet states for the same two nanodevices.
The results presented in Figs. 6 and 8
correspond to the nanodevice, in which the $y$ extension of the right QD is two times larger
than that of the left QD.  We see that the smaller values of $\varrho(x,0)$ in the larger
(right) QD [cf. Figs. 6(a,b)] are compensated by the larger spreading of the electrons in the $y$ direction
(cf. the right panel of Fig. 8).

\begin{figure}
\includegraphics[width=\linewidth,height=0.9\linewidth,angle=270]{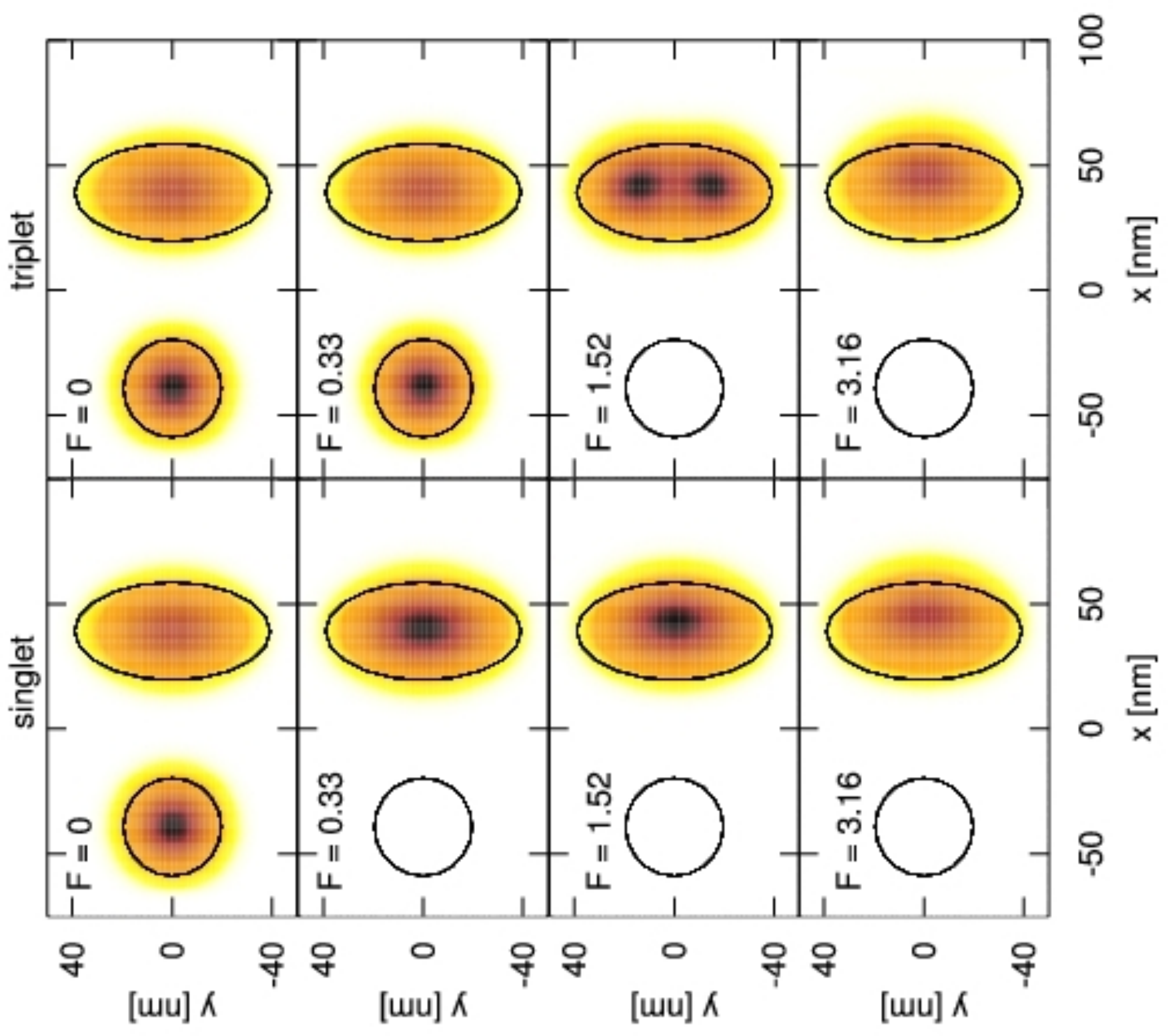}
\caption{\label{fig8} Contours of electron density for singlet (left panels)
and triplet (right panels) states on the $x-y$ plane for different values
of the external electric field $F$ (in kV/cm).
The darker color corresponds to the larger electron density.
Thin solid curves correspond to the sizes of the QD's.
The left QD has the circular shape with $R_{lx}=R_{ly} = 20$ nm and the right QD
has the elliptic shape with $R_{rx}=20$ nm, $R_{ry} = 40$ nm; the other parameters take on the values
 $p=10$ and $d = 80$ nm.}
\end{figure}

\begin{figure}
\includegraphics[width=0.9\linewidth,height=\linewidth,angle=270]{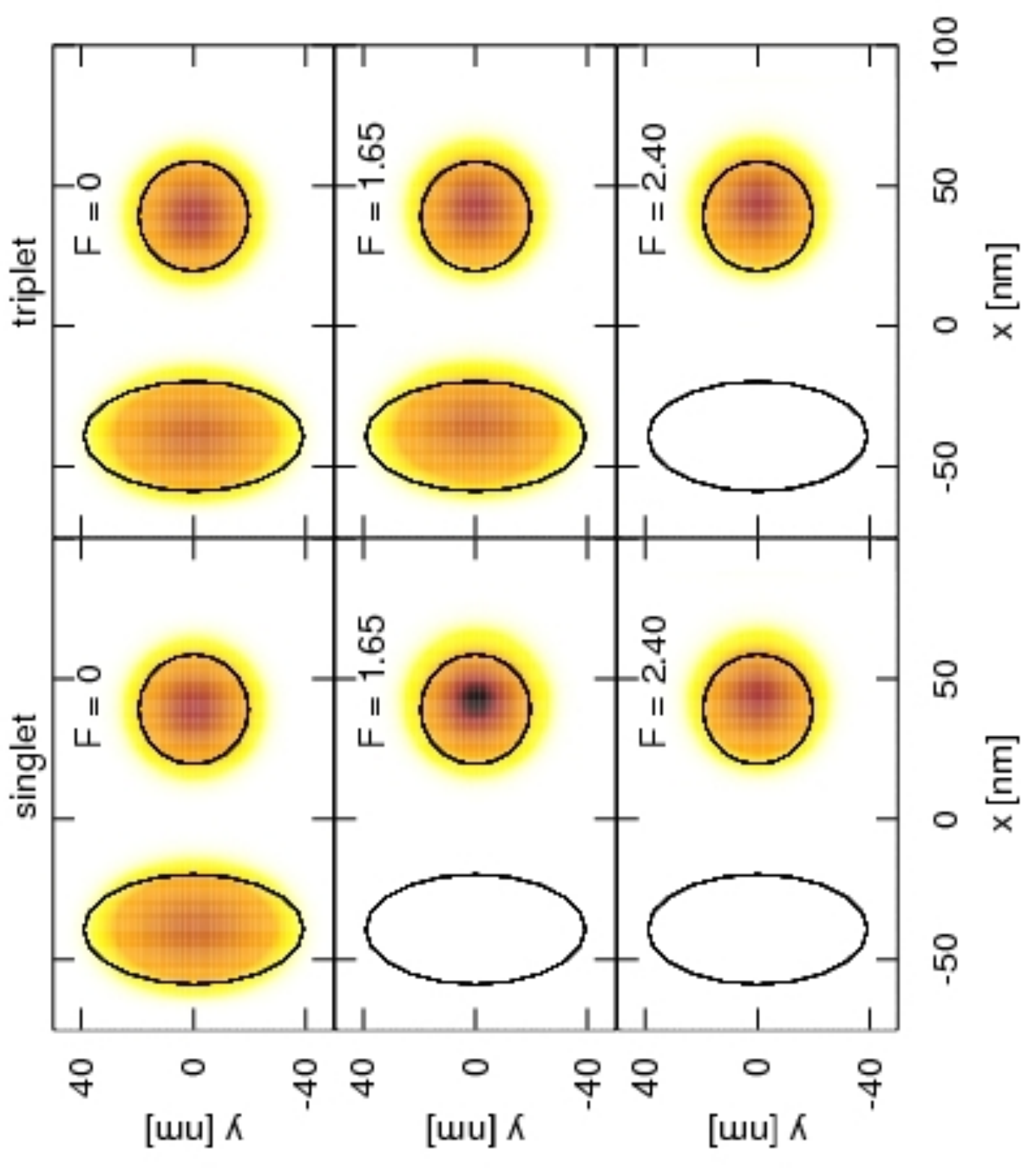}
\caption{\label{fig9}
Contours of electron density for singlet (left panels)
and triplet (right panels) states on the $x-y$ plane for different values
of the external electric field $F$ (in kV/cm).
The darker color corresponds to the larger electron density.
Thin solid curves correspond to the sizes of the QD's.
The left QD has the elliptic shape with
$R_{lx}=20$ nm and $R_{ly} = 40$ nm,
the right QD has the circular shape with $R_{rx}=R_{ry} = 20$ nm;
the other parameters take on the values $p=10$ and $d = 80$ nm.}
\end{figure}

The results displayed in Figs. 6-9 allow us to trace the changes of electron localization in
the two-electron system confined in the two coupled QD's, which result from the action
of the external electric field.
In the absence of the electric field, the electrons are localized in
the different QD's and there is no overlap between their wave functions
[cf. Figs. 6(a) and 7(a), and Figs. 8 and 9 for $F=0$].
Therefore, the exchange interaction between the electrons vanishes.
If we apply the external electric field,  the electrons start to tunnel
through the potential barrier from the left to the right QD and interact via the exchange
coupling [Figs. 6(b) and 7(b)].  In this regime, the increasing
electric field causes the fast linear increase of the exchange energy [Figs. 6(b) and 7(b)].
We note that -- even in the low electric field regime -- the electron distribution considerably changes
in the singlet state, but is only slightly distorted in the triplet
state (cf. Figs. 8 and 9 for $F=0$ and $F>0$).
If the right QD is sufficiently large,
the increasing electric field causes that both the electrons become localized
in this QD in either spin state [cf. Fig. 6(c) and Fig. 8 for
$F=1.52$ kV/cm].  In the singlet state, the electrons are localized
in the central region of the right QD, while in the triplet state, the electron density exhibits
two maxima clearly separated in the $y$ direction (cf. Fig. 8 for $F=1.52$ kV/cm).
In this field regime, the exchange energy reaches the largest values.
The double occupancy of the right QD and the separation of the triplet electrons in the $y$-direction
do not change in a rather broad range of the electric field,
which leads to the broad plateau region (Fig. 3).
If the electric field exceeds the critical value $F_{c2}$, one of the electrons tunnels through
the right triangular barrier out of the right QD and the exchange interaction rapidly falls down to zero [cf. Figs. 6 (d) and 8
for $F=3.16$ kV/cm and Figs. 7(d) and 9 for $F=2.40$ kV/cm].
The dotted (red) curves in the right part of Figs. 6(d) and 7(d) illustrate
the electron density just after the tunneling.
The tunneling electron is absorbed in the electron reservoir of the right electrode.
Let us note the corresponding changes of the one-electron energy levels (cf. the horizontal lines
in Figs. 6 and 7).  For $F \leq F_{c2}$ the one-electron energy levels lie below
the right-electrode continuum energy edge,
i.e., electrochemical potential $\mu_r = -eV$ of the right electrode.
This means that these electron states are bound.
For $F=F_{c2}$ the one-electron energy levels reach $\mu_r$, i.e.,
the electrons cease to be bound and form resonant states. Therefore, in the strong-field regime,
the electrons tunnel via these resonant states through
the right triangular barrier [Figs. 6(d) and 7(d)] to the right contact.

The nanodevice with the left QD larger than the right one (Figs. 7 and 9) shows
the similar electric-field behaviour to that obtained for the nanodevice with the left QD
smaller than the right one (Figs. 6 and 8) in the regime of low and intermediate electric fields.
We remind that -- in this regime -- the exchange energy is zero for $F \leq F_{c0}$ and grows linearly
with $F$ for $F_{c0} \leq F \leq F_{c1}$.  However, after reaching the maximum
value $J_{max}$ for $F=F_{c1}$, the exchange energy rapidly falls down to zero, i.e.,
the $J(F)$ dependence becomes qualitatively different from that shown in Figs. 3 and 6.
This behaviour results from the rapid change in electron localization
that occurs at $F=F_{c1}$.  For $F \leq F_{c1}$ the electrons in the triplet state are localized
in the different QD's [cf. Figs. 7(b,c)].  For the triplet state, the increase of the electric field above
$F_{c1}$ does not generate the double occupancy of the right QD, like in the nanodevice shown in Fig. 3,
but leads to the immediate tunneling of one of electrons to the right electron reservoir [Fig. 7(d)].
In the electric-field regime $F > F_{c1}$,
the right QD in this nanodevice can not be occupied by the two electrons in the triplet state,
which causes that the exchange interaction vanishes.
Fig. 7(d) shows that for the sufficiently strong electric field
the energy of the first excited one-electron state exceeds
the electrochemical potential of the right contact, i.e., the resonant state is formed.
In this case, we are dealing  with the resonant tunneling via the first excited one-electron state.
We note that this one-electron state yields the large contribution to the triplet two-electron wave function.

We have also studied the dependence of the exchange energy on the softness
of the confinement potential.  Figure 10 shows maximum exchange energy $J_{max}$
as a function of softness parameter $p$.  The maximal value $J_{max}$
is taken for $F=0.7576$ kV/cm, i.e., in the plateau region, for the nanodevice shown in the inset of Fig. 3.
We see that $J_{max}$ decreases if $p$ increases, i.e., if
the confinement potential becomes more hard.
This dependence can be approximated by the exponential function
\begin{equation}
J_{max}(p) = A_1 \exp(-C_1 p)+ B_1 \;,
\label{exp1}
\end{equation}
where $A_1=2.4740$ meV, $B_1=1.4059$ meV, and $C_1=0.2091$.
The exponential parametrization  [Eq.~(\ref{exp1})] results from the fact that --
for the fixed confinement-potential ranges --
the effective quantum capacity of the QD's increases with increasing $p$
(cf. inset of Fig. 10).
If $p$ increases, the electrons localized in the right QD become
more separated from each other, which leads to
the exponential decrease of the overlap of electron wave functions,
which in turn gives rise to the exponential decrease of the exchange energy.

\begin{figure}
\includegraphics[width=0.7\linewidth,height=\linewidth,angle=270]{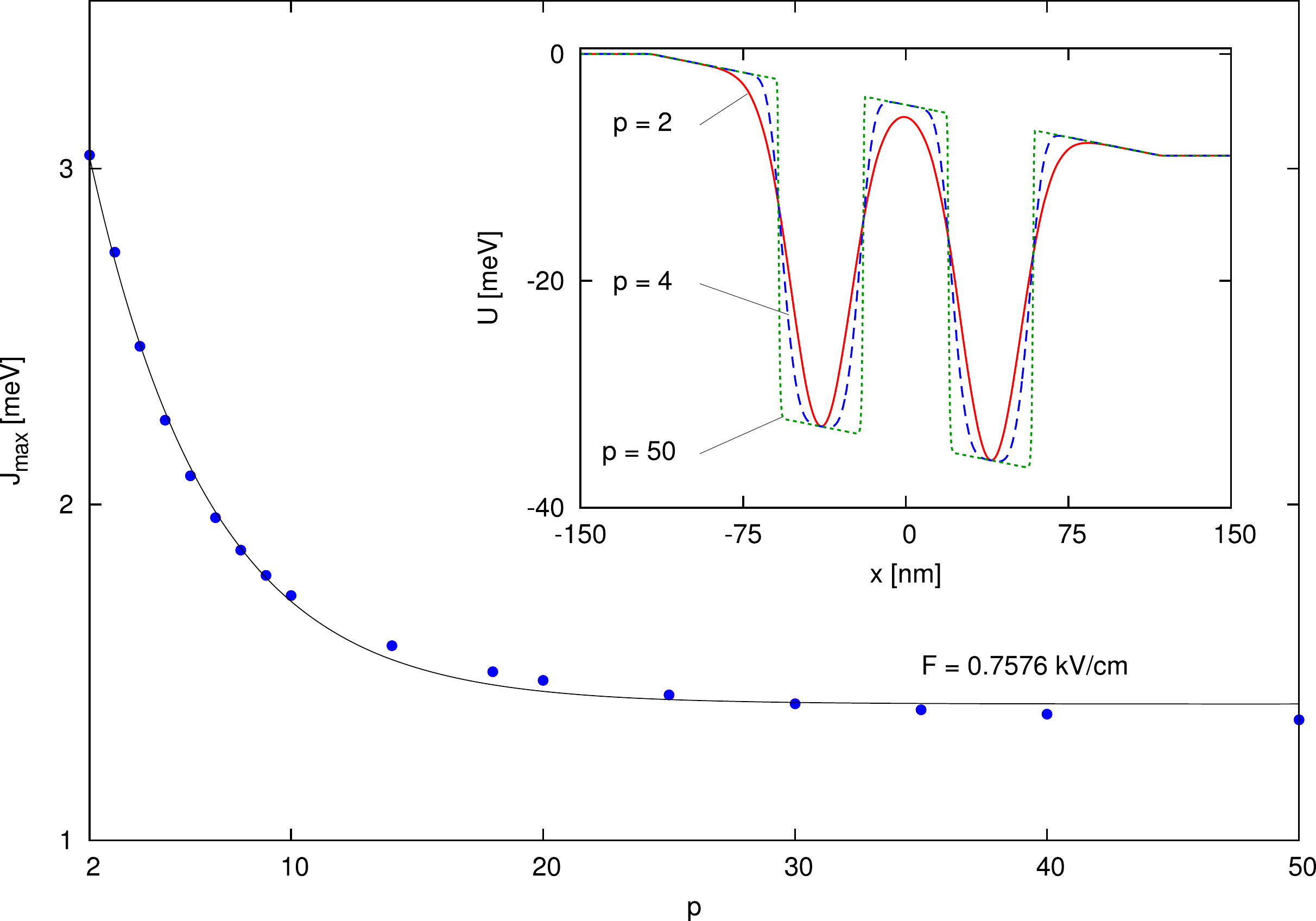}
\caption{\label{fig10} (Color online) Maximum exchange energy $J_{max}$
as a function of softness parameter $p$ for electric field $F=0.7576$ kV/cm.
Dots show the results of numerical calculations and the solid curve shows the fitted
exponential function [Eq.~(\ref{exp1})].
The parameters of the nanodevice: $R_{lx}=R_{ly} = 20$ nm, $R_{rx}=20$ nm, $R_{ry} = 40$ nm, and $d = 80$ nm.
Inset: Total potential energy $U$ of the electron as a function of $x$ and $p$ for $y=0$
and for fixed $F$ = 0.7576 kV/cm..}
\end{figure}

\begin{figure}
\includegraphics[width=0.7\linewidth,height=\linewidth,angle=270]{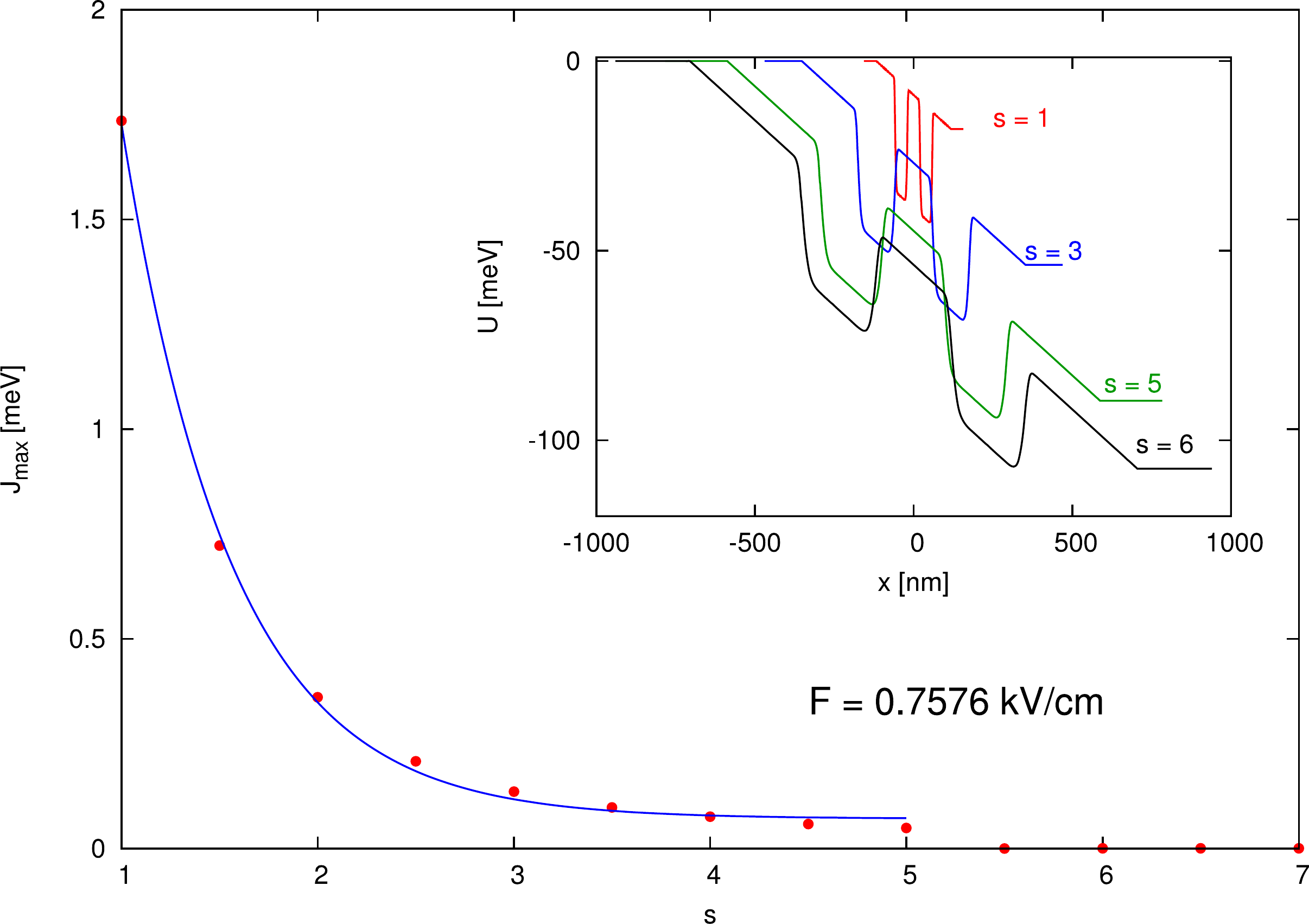}
\caption{\label{fig11} (Color online) Maximum exchange energy $J_{max}$
as a function of size scaling factor $s$ for $p=10$
and for electric field $F$ = 0.7576 kV/cm.
Dots show the results of numerical calculations and the solid curve shows the fitted
exponential function [Eq.~(\ref{exp2})].
The initial sizes ($s=1$) of the QD nanodevice are $R_{lx}=R_{ly} = 20$ nm, $R_{rx}=20$ nm,
$R_{ry} = 40$ nm, and $d = 80$ nm.
Inset: Total potential energy $U$ of the electron as a function of $x$ and $s$ for $y=0$
and for fixed $F$ = 0.7576 kV/cm.}
\end{figure}

For a possible experimental realization of the model nanodevices studied
in the present paper it is interesting to find a direct dependence of
the exchange energy on the size of the nanodevice.  For this purpose
we have calculated the maximum exchange energy
when scaling all the linear dimensions of the coupled QD system.
We consider the nanodevice displayed in the inset of Fig. 3,
for which the exchange energy takes on the maximal values in the broad plateau region
(cf. Fig. 3 for $p=10$).
We have defined the size scaling factor as $s=R_{actual}/R_{initial}$,
i.e., $s$ is equal to the ratio of the actual
linear dimension $R_{actual}$ to its initial value $R_{initial}$.
As the reference nanodevice with the initial values of the linear dimensions we take that
with $R_{lx}=R_{ly} = 20$ nm, $R_{rx}=20$ nm, $R_{ry} = 40$ nm, and $d = 80$ nm (cf. Fig. 3).
The calculations of $J_{max}$ have been performed for the set of
nanodevices characterized by $s$ times enlarged confinement potential ranges, i.e.,
$s R_{lx}, s R_{ly}, s R_{rx}$, and $s R_{ry}$ and interdot distance $sd$.  In the calculations,
we fix the strength $F$ of the electric field, i.e., we have to scale accordingly the interelectrode
distance $L \rightarrow sL$ and the applied voltage $V \rightarrow s V$.
The numerical results are displayed in Fig. 11 by the full (red) dots.
In the interval $1 \leq s \leq 5$  these results can be parametrized by the exponential function
\begin{equation}
J_{max}(s)= A_2 \exp(-C_2 s) + B_2 \; ,
\label{exp2}
\end{equation}
where $A_2$ = 9.9040 meV, $B_2 = 0.0710$ meV, and $C_2 = 1.7874$ (cf. solid curve in Fig. 11).
The exponential dependence [Eq.~(\ref{exp2})] is similar to that given by Eq.~(\ref{exp1}) and
can also be interpreted as resulting from the size effect.
If the total size of the nanodevice increases (cf. inset of Fig. 11),
the overlap between the one-electron wave function decreases exponentially with increasing $s$.
Also the localization of electrons in the QD's becomes weaker if the total size of the nanodevice grows.
We note that parametrization (\ref{exp2}) is valid for $s \leq 5$ only.  If the size of the nanodevice is
sufficiently large, i.e., the size scaling factor exceeds $\sim$ 5,
the exchange energy rapidly falls down to zero.  The disappearance of the exchange interaction
in the large-size nanodevice results from the delocalization of electrons, which
can be explained using the potential energy profiles (cf. inset of Fig. 11).
In order to keep the electric field constant when enlarging the nanodevice size $s$ times
we have to apply the $s$ times higher voltage.
This leads to the lowering of electrochemical potential $\mu_r=-eV$ of the right contact.
Simultaneously, the energy of the electrons localized in the right QD grows with respect to $\mu_r$.
We have checked that for $s \simeq 5$ the first excited-state one-electron energy level exceeds
the electrochemical potential of the right contact.
Moreover, the triangular barrier near the right contact
becomes more and more penetrable for the electrons if the size of the nanodevice increases.
In these conditions, one of the electron tunnels out of the right QD to the right reservoir
and the exchange interaction vanishes.

\section{Discussion}

The results of Figs. 3--5 show that -- in the nanodevice, which consists of the elliptic QD --
the static homogeneous electric field applied in the coupled QD's region
can switch on and off the exchange interaction.
In the nanodevice, which consists of the two circular QD's (Fig. 2),
the exchange interaction is non-zero at $F=0$ and
the electric field can only switch off the exchange interaction.
This behaviour (Fig. 2) is similar to that observed in the single QD.\cite{pss}
In the nanodevices shown in the insets of Figs. 2-4, the exchange energy exhibits a plateau in
a broad electric-field regime.  This plateau ends up at critical electric field
$F_{c2}$, above which the exchange energy rapidly vanishes.
If the electric field exceeds $F_{c2}$,
one of the triplet electrons tunnels from the right QD to the right contact
and the triplet state ceases to be bound.  Therefore, in this field regime, we can not speak
about the exchange interaction.  The electrons in the singlet state become unbound
if the electric field exceeds $F_{c2}$ by an amount $\Delta F_S$.
We have found that $\Delta F_S \simeq \Delta F_{linear}$,
i.e., $\Delta F_S$ is approximately equal to the width
of the linear $J(F)$ dependence [Eq.~(\ref{linear})].

The critical electric field $F_{c2}$ increases with increasing $p$ (cf. Figs. 2-4).
Simultaneously, the increasing $p$ leads to the decreasing maximum value
of the exchange energy reached in the plateau region (cf. Figs. 2 and 3).
Both these effects result from the increasing effective size of the QD's.
For fixed parameters $R_{lx}, R_{ly}, R_{rx}$, and $R_{ry}$
the effective size of the QD increases with increasing $p$, i.e., increasing hardness of the confinement
potential (cf. inset of Fig. 10).  This leads to the decreasing overlap between the electron wave functions
and the weaker electron localization, which in turn causes the decline
of the exchange energy.  Moreover, if the effective size of the QD's is larger,
we have to apply the stronger electric field in order to liberate
one of the electrons from the right QD, which gives rise to the increase of $F_{c2}$.

The critical electric field $F_{c0}$, below which $J=0$ and above which $J > 0$,
decreases with increasing $p$ (cf. Figs. 2-5).
This dependence results from the decreasing effective thickness of the potential barrier separating both the QD's
with increasing $p$ (cf. inset of Fig. 10).  If the barrier is thinner, the electrons tunnel through it with the larger probability
and the right QD becomes doubly occupied at the lower electric field.

In the low- and intermediate-field regime, the nanodevices with the laterally coupled QD's
possess an important property:
the exchange energy is a linear function of the electric field, i.e.,
it can be conveniently tuned by applying the external voltage.
In the nanodevice with the small right QD (cf. Fig. 5), the increasing electric field
switches on the exchange interaction at $F=F_{c0}$, leads to the linear increase of $J$
in a broad regime of $F$, and switches it off at $F=F_{c1}$.
Recently, the linear dependence of the exchange interaction energy on the electric field
has been found in vertically coupled self-assembled QD's.\cite{nowak}
In the present paper, we have obtained this linear dependence
for the laterally coupled QD's  with different geometric configurations
and different softness of the confinement potential (cf. Figs. 2-5).  Electric-field regime
$\Delta F_{linear}$ of this linear dependence is considerably broader in the nanodevices
that contain the small right QD (Fig. 5)
than in the nanodevices with the large right QD (Figs. 2-4).
According to Fig. 5, $\Delta F_{linear}$ extends to $\sim 0.8$ kV/cm.
For comparison, in Figs. 2-4, $\Delta F_{linear} \simeq$ 0.2 kV/cm.
The larger width of interval $\Delta F_{linear}$ obtained for the nanodevice depicted in the inset of Fig. 5
causes that the maximum value $J_{max}$ of the exchange energy is considerably larger than those for the nanodevices
shown in Figs. 2-4.  In the former case, $J_{max}$ reaches 6 meV.

\begin{figure}
\includegraphics[width=0.7\linewidth,height=\linewidth,angle=270]{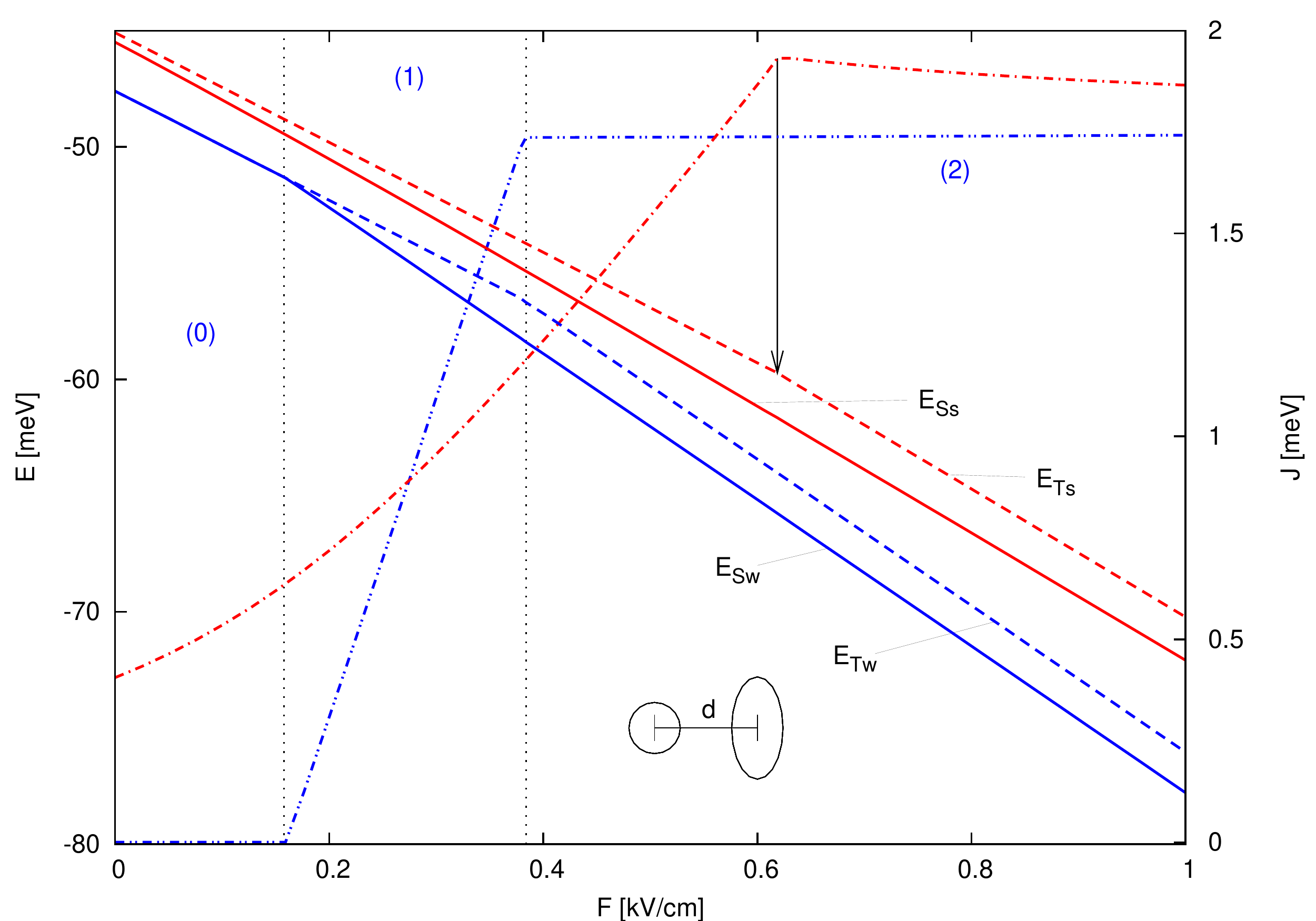}
\caption{\label{fig12} (Color online)
Energy of singlet ($E_{Sw}, E_{Ss}$) and triplet ($E_{Tw}, E_{Ts}$) states and exchange energy $J$
(dash-dotted curves)
as functions of electric field $F$ for two distances $d$ between the QD centers.
Inset shows the geometry of the nanodevice.
Subscripts $w$ and $s$ correspond to the weak ($d=80$ nm) and strong ($d=40$ nm) tunnel coupling,
respectively.  The other parameters of the QD's are the same as in Fig. 3 with $p=10$.
For clarity the curves with label $s$ are shifted upwards by 5 meV.
Thin vertical lines show critical fields $F_{c0}$ and $F_{c1}$ for the weak tunnel coupling,
the vertical arrow shows the cusp on the $E_{Ts}(F)$ plot, which gives rise to the
corresponding cusp on the $J(F)$ plot for the strong tunnel coupling.
Labels (0), (1), and (2) denote the electric-field regimes
$[0,F_{c0}], [F_{c0},F_{c1}]$, and $[F_{c1},F_{c2}]$, respectively,
for the weak tunnel coupling.}
\end{figure}

In order to get a more deep physical insight into the linear $J(F)$ dependence,
we have investigated the behaviour of the lowest singlet and triplet energy levels.
We have found that -- in the regime $0 \leq F  \leq F_{c2}$ --
the field dependence of these energy levels
can be very accurately parametrized by the linear functions (Fig. 12)
\begin{equation}
E_{S,T}(F)= \alpha_{S,T} F +\beta_{S,T} \;.
\label{STlinear}
\end{equation}
Parameters $\alpha_{S}$, $\alpha_{T}$, $\beta_{S}$, and $\beta_{T}$ take on different values in
the different regimes of the electric field,
i.e., $[0,F_{c0}], [F_{c0},F_{c1}], [F_{c1},F_{c2}]$,
that we label (0), (1), and (2), respectively, but are independent of $F$ within each regime.
In the low-field regime, i.e., for $0 \leq F \leq F_{c0}$,
$\alpha_S^{(0)} = \alpha_T^{(0)}$ and $\beta_S^{(0)} = \beta_T^{(0)}$, which leads to the zeroing of the
exchange energy, i.e., the singlet-triplet degeneracy.
In the intermediate-field regime, i.e., for $F_{c0} \leq F \leq F_{c1}$, the singlet-triplet degeneracy is lifted.
In this field regime, $\alpha_S^{(1)} < \alpha_T^{(1)}$ and $\beta_S^{(1)} > \beta_T^{(1)}$.
Therefore, we obtain $\alpha = \alpha_T^{(1)} - \alpha_S^{(1)} >0$
and $\beta = \beta_T^{(1)} - \beta_S^{(1)} <0$, which leads to the linear $J(F)$
dependence [Eq.~(\ref{linear})].

It is interesting that -- for the nanodevices shown in the insets of Figs. 2--3 --
linear parametrization (\ref{STlinear}) is also valid
in the regime of rather strong electric fields, i.e., for $F_{c1} \leq F \leq F_{c2}$.
The values of parameters $\alpha_{S,T}^{(2)}$ and $\beta_{S,T}^{(2)}$ are different from
those obtained for the lower fields, but are approximately constant within this field regime.
In field regime (2), $\alpha_S^{(2)} \simeq \alpha_T^{(2)}$ and $\beta_T^{(2)} > \beta_S^{(2)}$,
which gives rise to the plateau of the exchange energy (cf. Figs. 2--4)
with $J_{max} \simeq \beta_T^{(2)} - \beta_S^{(2)} = const$.

We have also found another interesting feature of the linear parametrization (\ref{linear}).
Parameter $\alpha=\Delta J/\Delta F$ that determines the rate of changes of the exchange energy with the electric field
in the linear regime takes on almost the same values for all the nanodevices described
by the parameters quoted in the captions of Figs. 2-5.
Parameter $\alpha$ is independent of the softness of the confinement potential
and the geometry of the nanodevice (cf.  Figs. 2-5).
Considering all the $J(F)$ dependencies, displayed in Figs. 2-5, i.e.,
studying several different nanodevices, we have found that $\alpha = 7.73 \pm 0.13$ [meV/(kV/cm)].
In the lateral QD's, the linear $J(F)$ dependence is a non-trivial property that can not
be explained by the non-degenerate first order perturbation theory.\cite{nowak}  This effect occurs
in the intermediate-field regime, in which the electron wave functions are considerably distorted
with respect to those for $F=0$ (cf. Figs. 6-9).
In particular, we note the rapid change of the singlet-state localization
in the low- and intermediate-field regime (Figs. 8 and 9).
Moreover, we have found that even for the same geometric configuration of the QD's and in the same electric-field regime,
the linear $J(F)$ dependence disappears if the tunnel coupling between the QD's is sufficiently strong.
In order to show this effect, we have considered the two nanodevices with the geometric configuration shown in the insets
of Figs. 3 and 12.
Fig. 12 displays the results for the two nanodevices characterized by different separations $d$ between the
QD centers and the same values of all other parameters.  For $d=80$ nm, i.e., for the weak interdot tunnel coupling,
the singlet and triplet energies as well as the exchange energy are linear functions of
the electric field.  We also observe the cusp on the curve $E_{Tw}(F)$, which leads to
the corresponding cusp on the curve $J(F)$.
In the low electric-field regime and for the weak tunnel coupling,
the exchange interaction vanishes due to the singlet-triplet degeneracy.
For $d=40$ nm, i.e., for the strong tunnel coupling,
a completely different $J(F)$ behaviour has been obtained.  In this case, the singlet-state energy
is a non-linear function of the electric field, which leads to the non-linear $J(F)$ dependence.
However, the triplet energy $E_{Ts}$ is a piece-wise linear function of $F$
with the cusp shown by the arrow in Fig. 12.
In the low electric-field regime and for the strong tunnel coupling, the singlet-triplet degeneracy is lifted
and the exchange interaction is non-zero even for $F=0$.

\begin{figure}
\includegraphics[width=0.8\linewidth,height=\linewidth,angle=270]{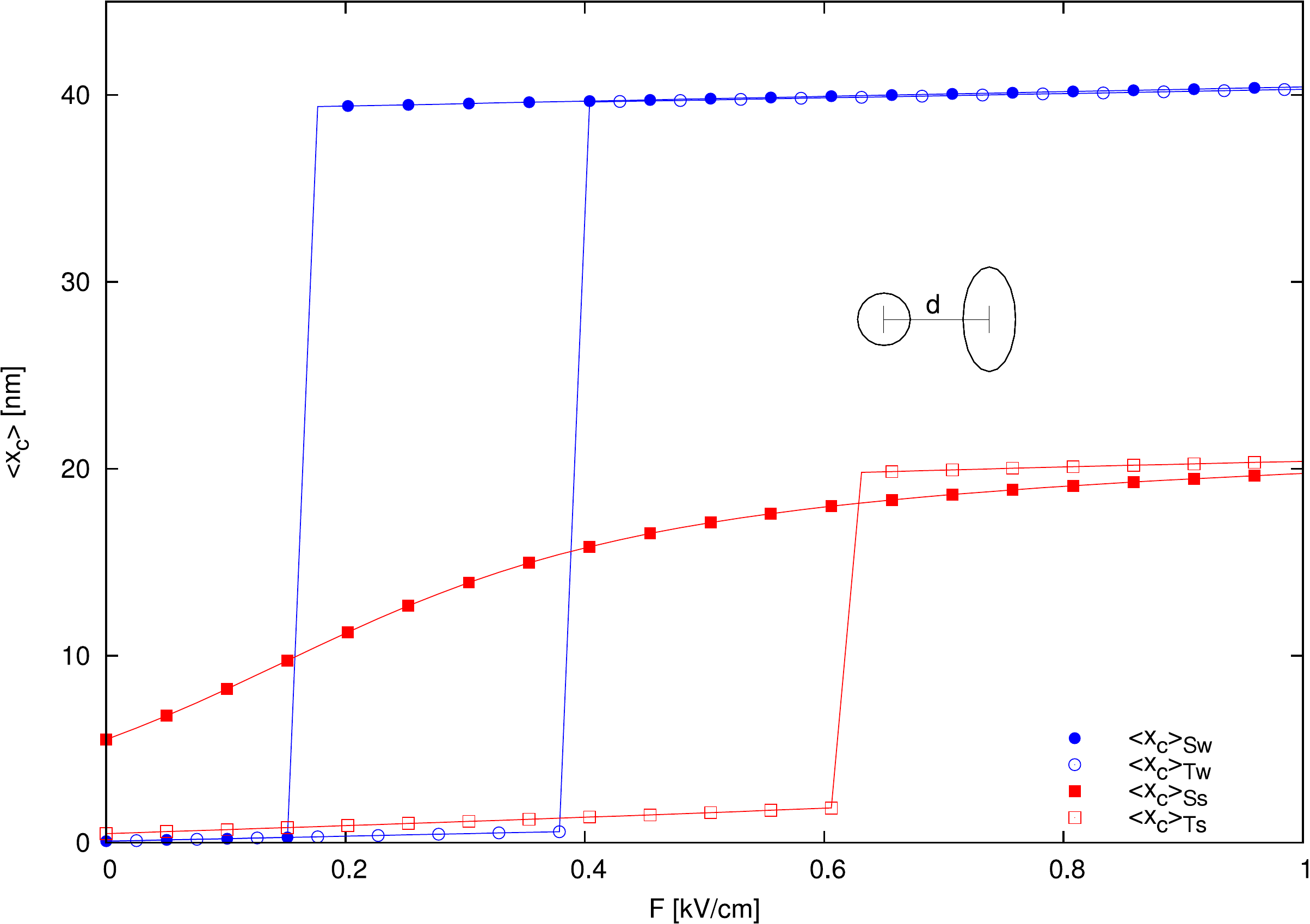}
\caption{\label{fig13} (Color online)
Expectation value $\langle x_c \rangle$ of the charge gravity center position as a function of electric field $F$
for the singlet (subscript $S$, curves with full circles and squares) and triplet states
(subscript $T$, curves with open circles and squares) for $d=40$ nm (subscript $s$, red curves)
and $d=80$ nm (subscript $w$, blue curves).  The nanodevice is the same as in Fig. 12.}
\end{figure}

The linear/non-linear $J(F)$ dependence can be further discussed if we consider
the changes of the charge gravity center position, i.e., $x_c=(x_1+x_2)/2$.
We have calculated the expectation value $\langle x_c \rangle$
for the weak and strong interdot tunnel coupling using the field-dependent two-electron wave functions
$\Psi(\mathbf{r}_1,\mathbf{r}_2)$ for the singlet and triplet states.
According to Eq.~(\ref{DU}),  the value 2$\langle x_c \rangle$ determines the electric-field
contribution to the potential energy of two electrons.  Figure 13 shows that
$\langle x_c \rangle$ is a linear function of $F$ for the triplet states in both the cases of the weak and strong
tunnel coupling and for the singlet state only in the case of the weak tunnel coupling.
However, for the singlet state and the strong tunnel coupling the dependence of
$\langle x_c \rangle_{Ss}$ on the electric field is nonlinear.
We also observe that the expectation values $\langle x_c \rangle_{Sw}$, $\langle x_c \rangle_{Tw}$,
and $\langle x_c \rangle_{Ts}$ exhibit jumps at certain values of the electric field.
Besides they are linear functions of $F$.
Figure 13 shows that the charge gravity center position follows the electron distribution
shown in Fig. 8 for $d=80$ nm.  At low electric fields, the center of charge gravity is localized
near the center of the nanodevice, i.e., at $x \simeq 0$,
for the triplet as well as the singlet states,
which results from the single occupancy of both the QD's.
The increasing electric field only slightly shifts
$\langle x_c \rangle$ to the right QD.  The single QD occupancy remains unchanged
for the triplet state up to $F=F_{c1}$.  However,  for the singlet state
the second electron tunnels to the right QD at $F=F_{c0}$,
which leads to the jump of $\langle x_c \rangle_{Sw}$  (Fig. 13).
For $F \geq F_{c1}$ and the weak tunnel coupling the charge gravity center
is approximately localized near the center of the right QD in both the spin states
(cf. the plots of $\langle x_c \rangle_{Sw}$ and $\langle x_c \rangle_{Tw}$ in Fig. 13).
For the weak tunnel coupling, the localization of the electrons in the singlet state
remains almost unchanged in the regime $F_{c0} \leq F \leq F_{c1}$.
For the strong tunnel coupling $\langle x_c \rangle_{Ss}$ in the singlet state
increases as a non-linear function of $F$ starting from a non-zero
value for $F=0$.
In the triplet state, the jump of $\langle x_c \rangle_{Ts}$ at $F=F_{c1}$ originates from
the rapid change of the localization from the single to double occupancy of the right QD.

The results of Fig. 13 can be translated into the energy dependencies displayed in Fig. 12.
The linear $J(F)$ dependence occurs if the interdot tunnel coupling is weak.
Then, in the triplet state, the single-electron occupancy of each QD occurs
in electric-field regime $[0,F_{c1}]$.
In the field regime $[F_{c0},F_{c1}]$, the right QD is doubly occupied
by the singlet electrons.  If the interdot separation is sufficiently large,
the increasing electric field slowly shifts the center of the charge gravity towards the right QD.
Simultaneously, the electrons are compressed at the right potential barriers
of QD's, which hampers this shift.
The net shift is proportional to the electric field,
which results in the linear dependencies $E_S(F)$ and $E_T(F)$.
If the interdot tunnel coupling is sufficiently strong, the electrons can tunnel through
the barrier with a rather large probability, which
results from the considerable overlap of the wave functions centered in both the QD's.
Therefore, for the strong tunnel coupling the increasing electric field leads to
the non-linear increase of the exchange energy.

\section{Conclusions and Summary}

The results of the present paper allow us to discuss the effect of external electric fields
of different origin on the electronic properties of QD nanodevices.
The electric fields can be created by the different electrodes, which surround the QD region.
In a direct way, we have investigated the effect of the static homogeneous electric field,
which is usually created by the source and drain electrodes.
In an indirect way, we have also studied the effect of the gate electrodes.
We are able to determine the effect of the gates, since the gate-controlled QD's
(electrostatic QD's) \cite{hand,lis03,lis08}
are induced by the inhomogeneous electric field, which is created by the gates.
Therefore, the QD confinement potential, in particular, its shape, range, potential-well depth,
and softness, are determined by the voltages applied to the gates.
This means that studying the dependence of the electron states on these parameters
of the confinement potential we are investigating the effect of the gates.

In the present paper, we focus on the exchange interaction, which plays an important role
in a manipulation with spin qubits.\cite{loss98,burk99,burk00,mosk07}
The exchange energy can be effectively tuned by changing
the external electric field, i.e., changing the voltages applied to the electrodes.
By increasing the homogeneous electric field (bias voltage) we can switch on/off the exchange interaction.
The critical values of the electric field, for which this on/off switching occurs,
depend on the softness of the confinement potential, i.e., can be changed by changing the gate voltages.
We have found the existence of the plateau in the exchange energy versus electric field dependence.
This plateau occurs if the QD, into which the electrons are pushed by the electric field,
is larger than the other QD.
In the plateau regime, the exchange energy take on is maximal values.
Therefore, the on/off switching the exchange interaction occurs between zero and maximum value.
We have determined the critical electric fields and optimal nanodevice parameters, for which
this switching is the most effective.

We have shown that -- in the nanodevices with the weakly coupled lateral QD's at moderate
electric fields -- the exchange energy is a linear function of the electric field.
We have found that the parameter $\alpha=\Delta J/\Delta F$ that determines
the rate of changes of $J(F)$ is nearly the same for different nanodevices.
The constancy of $\alpha$ suggests that this parameter is universal
for a large class of nanodevices based on the laterally coupled QD's
provided that the interdot tunnel coupling is weak.
We have also demonstrated that for the sufficiently strong interdot tunnel coupling
the $J(F)$ dependence becomes non-linear.

The numerical approach proposed (see Appendix) is convenient for the calculations
of few-electron states in QD's in the external electric field.
Among several advantages (cf.  Appendix) of this approach,
we would like to underly the most important one.
Namely, this {\em numerical} procedure possesses the following {\em physical} property:
in the low- and intermediate-field regime,
it allows us to describe bound states of electrons in the electric field of finite range,
which acts in the real nanodevices.  Let us mention that in the electric field of infinite range,
usually assumed in the papers on this subject, \cite{zh06,nowak}
we always deal with non-stationary
states that -- for weak fields only -- can be treated as quasi-bound.

Having at disposal the nanodevices based on the gate-controlled lateral QD's,
we can intentionally change the shape, range, and potential-well depth
of the confinement potential by changing the voltages applied to the gates.
The manipulation of electronic states in laterally coupled gate-controlled QD's,
performed by changing the gate voltages, is very effective.
Therefore, the nanodevices based on the lateral QD's
are very promising in quantum computing based on spin qubits.
In the present paper, we have shown how to control and tune
the exchange interaction by changing the external electric field.
Our results should be helpful in designing the nanodevices for the spin qubit processing
with the help of the controlled exchange interaction.

\acknowledgments

This paper has been supported in part by the Polish Ministry of
Science and High School Education in the frame of grant No. N N202 173835.

\appendix*
\section{Computational method}

We give a brief description of the numerical method proposed in the present paper in order to
obtain the one-electron wave functions in the discrete representation
that is convenient in the two-electron calculations.
First, we set up the computational box for given geometric parameters of the QD nanodevice.
Usually, we take on its $x$ extension $L_x=400$ nm  and $y$ extension $L_y=140$ nm.
Next, we solve the one-electron Schr\"{o}dinger equation by variational means.
For this purpose we apply the multi-center Gaussian basis defined
inside the computational box
\begin{equation}
g_{ij}(x,y)=\exp[-\gamma_1(x-x_i)^2-\gamma_2(y-y_j)^2] \;,
\label{Gauss}
\end{equation}
where $\gamma_1$ and $\gamma_2$ are the variational parameters.

\begin{figure}
\includegraphics[width=\linewidth,height=\linewidth,angle=0]{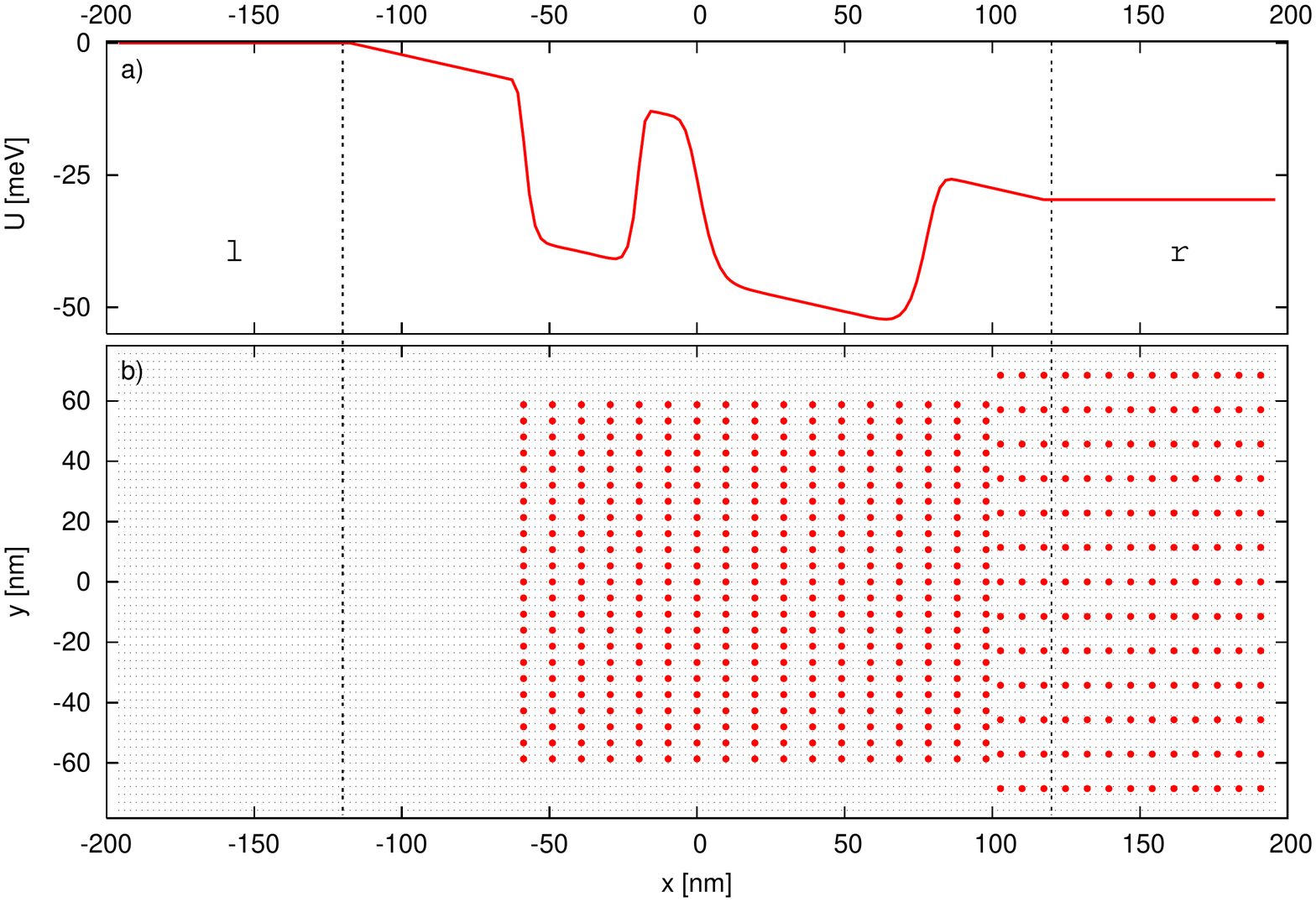}
\caption{\label{fig14} (Color online) (a) Profile of electron potential energy $U$
as a function of $x$ for $y=0$ and $F = 1.2625$ kV/cm.
Symbols $l$ and $r$ denote the left and right electrode, respectively.
Vertical dashed lines correspond to the boundaries of the electrodes.
(b)  Schematic of the computational box.  Thick (red) dots  display the positions of the centers of Gaussians
and thin (black) dots show the grid points used in discrete representation (\ref{discr}).
The nanodevice parameters used in the plots: $R_{lx}=R_{ly} = 20$ nm, $R_{rx}=40$ nm, $R_{ry} = 40$ nm, $p=10$,
 and $d = 80$ nm.}
\end{figure}

The centers of Gaussians $(x_i,y_j)$, where $i=1, \ldots , I$ and $j=1, \ldots, J$,
form a grid within a large rectangle,
which encompasses both the QD's. i.e., the edges of the rectangle are longer than
the  double range of the confinement potential in each direction (cf. Fig. 14).
Gaussians (\ref{Gauss}) are chosen to cover the region of electron localization in the QD's
and the region of the right electrode, to which the electrons are shifted.
The variational wave function $\phi_{\nu}(x,y)$ for one-electron state $\nu$ is taken on in the form of
a linear combination of Gaussians (\ref{Gauss})
\begin{equation}
\phi_{\nu}(x,y) = C_{\nu}\sum\limits_{i=1}^{I}\sum\limits_{j=1}^J
c^{\nu}_{ij}g_{ij}(x,y) \;,
\label{var}
\end{equation}
where $C_{\nu}$ is the normalization constant.
The values of $I$ and $J$ are chosen according to the actual size of the nanodevice
and extend up to $I_{max}=50$ and $J_{max}=33$.
This means that we have at disposal 1650 basis elements
(\ref{Gauss}), which assures the high accuracy of one-electron solutions.
The matrix elements of one-electron Hamiltonian (\ref{h1}) are calculated in basis (\ref{Gauss})
as follows: the matrix elements of the kinetic energy are calculated analytically
and the matrix elements of the potential energy
are calculated by a numerical quadrature.
Solving the generalized eigenvalue problem of Hamiltonian (\ref{h1}) in basis (\ref{Gauss})
we obtain linear parameters $c^{\nu}_{ij}$ and energy eigenvalues $E_{\nu}$ for one-electron states
$\nu=1, \ldots , N_{\nu}$.
The values of non-linear variational parameters $\gamma_1$ and $\gamma_2$
are determined from the minimization of the ground-state energy $E_0$.
We have checked that $\gamma_1$ and $\gamma_2$ change only slightly when minimizing the excited-state
energy levels; therefore, we take $\gamma_1$ and $\gamma_2$ to be the same for each state $\nu$.

In the last step, we define the fine grid $(x_m,y_n)$, where $m=1, \ldots , M$ and $n=1, \ldots , N$,
and find the discrete representation of the one-electron wave functions by setting the $M \times N$ matrix
\begin{equation}
\phi_{\nu}^{mn} = \phi_{\nu}(x_m, y_n) \;.
\label{discr}
\end{equation}
The one-electron wave functions in representation (\ref{discr}) are used to construct the Slater determinants
and next to perform the CI calculations.
In the two-electron calculations, we have used $N_{\nu}=13$ one-electron states to construct up to
$N_S=169$ Slater determinants.  The number of mesh points was $M\times N=123\times 91$.

We would like to emphasize the following advantages of the present approach:
(i) It allows us to solve the electron eigenproblem in the real nanodevice, in which
the electric field has the finite range, i.e., the voltage is applied between the electrodes
separated by the finite distance.
This numerical feature possesses an important physical consequence:
if the electric field is not too strong, we are dealing with the well-defined stationary bound states.
It is commonly assumed that the electric field possesses the infinite range,
i.e., instead of the form (\ref{DU}) the potential energy has the form
$\Delta U = -eFx$ for $-\infty < x < +\infty$.  In this case, the electron states
are unbound for arbitrary $F$. This assumption leads to serious problems in the physical interpretation and
does not allow us to describe rigorously the real finite-size nanodevices.
(ii) It can be applied to arbitrary confinement potential
and inhomogeneous electric field.  In particular, this approach can be easily extended
to multiple coupled QD's with more than two QD's.
(iii) The present approach can be extended to few-electron systems with more than two electrons
and to three-dimensional (3D) nanostructures.

\end{document}